\documentclass{nature}
\usepackage{graphicx,psfrag}
\usepackage{mathrsfs}
\usepackage{amsmath,amsfonts,amssymb}
\usepackage{multirow}
\usepackage{comment}
\usepackage{units}
\usepackage{enumerate}
\usepackage[normalem]{ulem} 
\usepackage{longtable}
\usepackage[rgb]{xcolor}
\usepackage{gensymb}
\usepackage{adjustbox}
\usepackage{geometry}
\usepackage{array}
\usepackage[none]{hyphenat}
\usepackage{multibib}
\newcites{main}{References}
\newcites{methods}{References (Methods)}
\usepackage[colorlinks=true, urlcolor=blue]{hyperref}

\newcommand\blfootnote[1]{%
	\begingroup
	\renewcommand\thefootnote{}\footnote{#1}%
	\addtocounter{footnote}{-1}%
	\endgroup
}

\newcommand{\customfootnote}[2]{{
		\renewcommand{\thefootnote}{#1}
		\footnote[0]{#2}}}

\title{Bright X-ray and Radio Pulses from a Recently Reactivated Magnetar}

\newcommand{\Pearlman}{Aaron~B.~Pearlman}
\newcommand{\Majid}{Walid~A.~Majid}
\newcommand{\Prince}{Thomas~A.~Prince}
\newcommand{\Ray}{Paul~S.~Ray}
\newcommand{\Kocz}{Jonathon~Kocz}
\newcommand{\Horiuchi}{Shinji~Horiuchi}
\newcommand{\Naudet}{Charles~J.~Naudet}
\newcommand{\Guver}{Tolga~G\"uver}
\newcommand{\Enoto}{Teruaki~Enoto}
\newcommand{\Arzoumanian}{Zaven~Arzoumanian}
\newcommand{\Gendreau}{Keith~C.~Gendreau}
\newcommand{\Ho}{Wynn~C.~G.~Ho}

\author{\Pearlman$^{1,11,12,*}$\blfootnote{\textsuperscript{11}~NDSEG Research Fellow.}\blfootnote{\textsuperscript{12}~NSF Graduate Research Fellow.}\blfootnote{\textsuperscript{*}~Corresponding author: Aaron~B.~Pearlman~\textcolor{blue}{(aaron.b.pearlman@caltech.edu)}}, \Majid$^{2,1}$, \Prince$^{1,2}$, \Ray$^{3}$, \Kocz$^{1,4}$, \Horiuchi$^{5}$, \Naudet$^{2}$, \Enoto$^{6}$, \Guver$^{7,8}$, \Arzoumanian$^{9}$, \Gendreau$^{9}$, \Ho$^{10}$}

\newcommand{\CaltechPhysics}{Division of Physics, Mathematics, and Astronomy, California Institute of Technology, Pasadena, CA 91125, USA}
\newcommand{\JPL}{Jet Propulsion Laboratory, California Institute of Technology, Pasadena, CA 91109, USA}
\newcommand{\NRL}{Space Science Division, U.S. Naval Research Laboratory, Washington, DC 20375, USA}
\newcommand{\Berkeley}{Department of Astronomy, University of California, Berkeley, CA 94720, USA}
\newcommand{\CDSCC}{CSIRO Astronomy and Space Science, Canberra Deep Space Communications Complex, P.~O.~Box~1035, Tuggeranong, ACT~2901, Australia}
\newcommand{\RIKEN}{Extreme Natural Phenomena RIKEN Hakubi Research Team, RIKEN Cluster for Pioneering Research, 2-1 Hirosawa, Wako, Saitama 351-0198, Japan}
\newcommand{\ASSIST}{Department of Astronomy and Space Sciences, Faculty of Science, Istanbul University, Beyaz\i t, 34119 Istanbul, Turkey}
\newcommand{\IUORAC}{Istanbul University Observatory Research and Application Center, Beyaz\i t, 34119 Istanbul, Turkey}
\newcommand{\GSFC}{X-ray Astrophysics Laboratory, NASA Goddard Space Flight Center, Greenbelt, MD 20771, USA}
\newcommand{\Haverford}{Department of Physics and Astronomy, Haverford College, 370 Lancaster Avenue, Haverford, PA 19041, USA}

\begin{document}

\hypersetup{
	colorlinks,
	linkcolor={blue!50!black},
	citecolor={blue!50!black},
	urlcolor={blue!50!black}
}

\newgeometry{top=2.5cm}

\maketitle

\begin{affiliations}
	\item~\CaltechPhysics
	\item~\JPL
	\item~\NRL
	\item~\Berkeley
	\item~\CDSCC
	\item~\RIKEN
	\item~\ASSIST
	\item~\IUORAC
	\item~\GSFC
	\item~\Haverford
\end{affiliations}

\begin{abstract}

Magnetars are young, rotating neutron stars that possess larger magnetic fields \\ ($B$\,$\approx$\,10$^{\text{13}}$--10$^{\text{15}}$\,G) and longer rotational periods ($P$\,$\approx$\,1--12\,s) than ordinary pulsars\citemain{Olausen+2014, Kaspi+2017}. In contrast to rotation-powered pulsars, magnetar emission is thought to be fueled by the evolution and decay of their powerful magnetic fields. They display highly variable radio and X-ray emission\citemain{Camilo+2006, Camilo+2007b, Pearlman+2018, Levin+2010}, but the processes responsible for this behavior remain a mystery. We report the discovery of bright, persistent individual X-ray pulses from XTE~J1810--197, a transient radio magnetar, using the \textit{Neutron star Interior Composition Explorer~(NICER)} following its recent radio reactivation\citemain{Lyne+2018}. Similar behavior has only been previously observed from a magnetar during short time periods following a giant flare\citemain{Hurley+1999, Palmer+2005}. However, the X-ray pulses presented here were detected outside of a flaring state. They are less energetic and display temporal structure that differs from the impulsive X-ray events previously observed from the magnetar class, such as giant flares\citemain{Hurley+1999, Palmer+2005} and short X-ray bursts\citemain{Woods+2005}. Our high frequency radio observations of the magnetar, carried out simultaneously with the X-ray observations, demonstrate that the relative alignment between the X-ray and radio pulses varies on rotational timescales. No correlation was found between the amplitudes or temporal structure of the X-ray and radio pulses. The magnetar's 8.3\,GHz radio pulses displayed frequency structure, which was not observed in the pulses detected simultaneously at 31.9\,GHz. Many of the radio pulses were also not detected simultaneously at both frequencies, which indicates that the underlying emission mechanism producing these pulses is not broadband. We find that the radio pulses from XTE~J1810--197 share similar characteristics to radio bursts detected from fast radio burst~(FRB) sources, some of which are now thought to be produced by active magnetars\citemain{Bochenek+2020b}.

\end{abstract}

XTE~J1810--197 was discovered with the \textit{Rossi X-ray Timing Explorer~(RXTE)} during an X-ray outburst that began in 2003\citemain{Ibrahim+2004} and lasted until early 2007\citemain{Camilo+2016}. It is located at a distance of 3.1--4.0\,kpc\citemain{Minter+2008}, making it one of the nearest magnetars. The pulsar has a rotational period of 5.54\,s and a soft X-ray spectrum\citemain{Ibrahim+2004}. Highly linearly polarized radio pulsations were first detected from the magnetar in 2006 using the Parkes telescope, and bright, narrow radio single pulses were observed during each rotation of the neutron star\citemain{Camilo+2006}. This discovery established that a relationship exists between magnetars and the larger population of ordinary radio pulsars. XTE~J1810--197 has a dispersion measure~(DM) of 178\,$\pm$\,5\,pc\,cm$^{\text{--3}}$ and a spectral index of --0.5\,$\lesssim$\,$\alpha$\,$\lesssim$\,0 between 1.4 and 144\,GHz\citemain{Camilo+2006, Camilo+2007c}, with a radio flux density given by $S_{\nu}$\,$\propto$\,$\nu^{\alpha}$. The magnetar's flat radio spectrum has enabled the detection of pulsed radio emission at much higher frequencies than is typically observed from most radio pulsars.

In late~2008, radio pulsations from XTE~J1810--197 suddenly ceased\citemain{Camilo+2016}, and the magnetar remained in a quiescent state for more than a decade\citemain{Pintore+2019}. However, on 2018 December~8 (MJD\,58460), radio pulsations were redetected from the magnetar using the 76\,m Lovell Telescope at Jodrell Bank Observatory\citemain{Lyne+2018}. Following the magnetar's reactivation, X-ray and radio follow-up observations were performed (e.g. see refs.\,\citemain{Majid+2019, Dai+2019, Gotthelf+2019, Guver+2019}).

We carried out observations of XTE~J1810--197 with the X-ray Timing Instrument~(XTI) on board \textit{NICER} between 2019 February~6 (MJD\,58520) and 2019 February~26 (MJD\,58540). High frequency radio observations of the magnetar were also performed simultaneously at 8.3 and 31.9\,GHz using the NASA Deep Space Network~(DSN)\citemain{Pearlman+2019} 34\,m radio telescopes near Canberra, Australia on 2019 February~16 (MJD\,58530) and 2019 February~25 (MJD 58539), which included times when \textit{NICER} was also observing the source. These instruments and the data reduction procedures are described in Sections~\ref{Section:XrayObservations} and~\ref{Section:RadioObservations} (Methods). A catalog of the X-ray and radio observations presented in this Letter is provided in Table~\ref{Table:Table1}. Measurements of the magnetar's mean flux density and spectral index at 8.3 and 31.9\,GHz are listed in Table~\ref{Table:Table2}. Unless otherwise stated, all errors quoted in this paper correspond to 1$\sigma$ uncertainties.

\begin{figure}
	\centering
	\includegraphics[trim=0cm 0cm 0cm 0cm, clip=false, scale=0.4, angle=0]{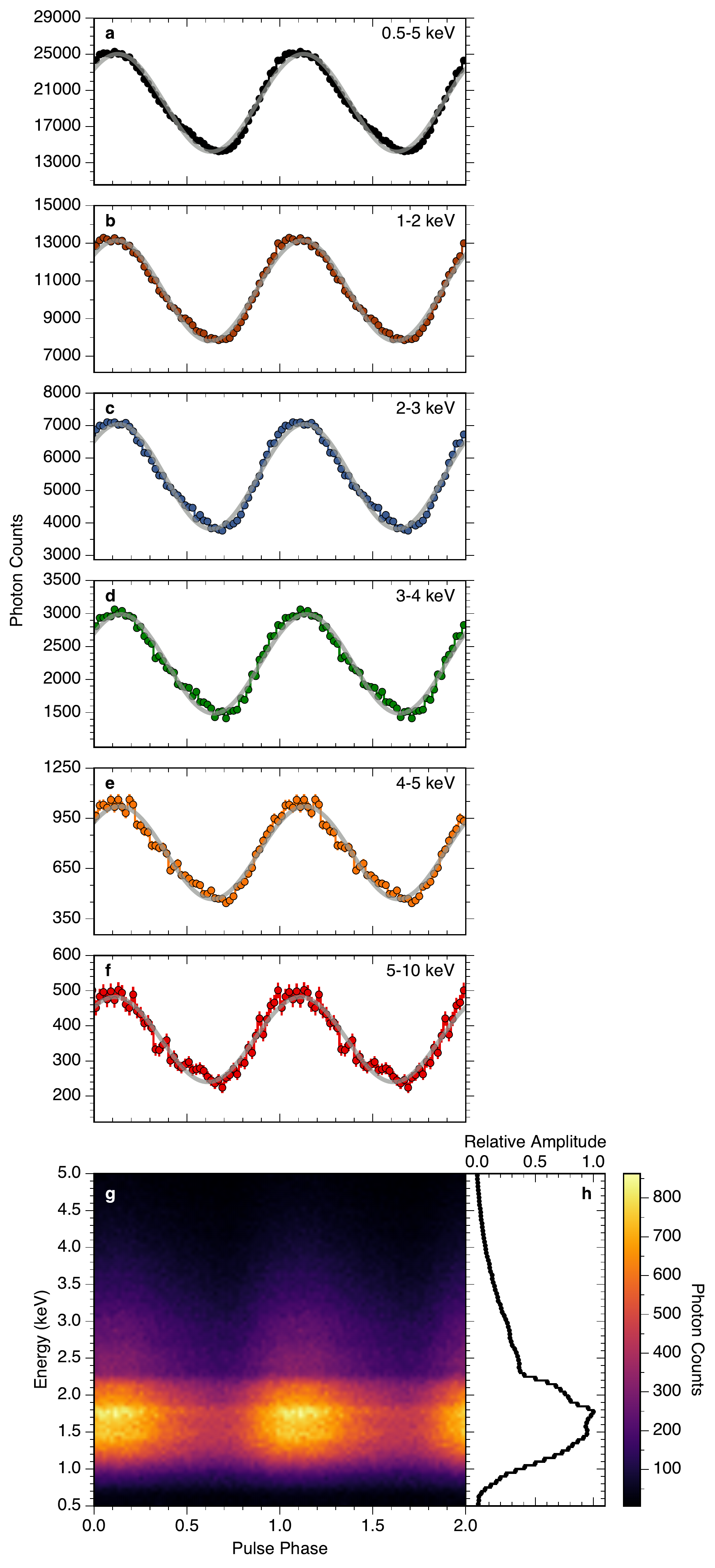}
	\caption{X-ray pulse profiles of XTE~J1810--197 in the (a)~0.5--5, (b)~1--2, (c)~2--3, (d)~3--4, (e)~4--5, and (f)~5--10\,keV energy bands. The pulse profiles are derived by combining all of the data from the \textit{NICER} observations listed in Table~\ref{Table:Table1}. Each pulse profile is folded with 50 phase bins using an ephemeris derived from radio pulsar timing measurements between MJDs\,58521 and 58540, where phase~0 corresponds to MJD\,58530.761334907~(TDB). Best-fit sinusoids to the pulse profiles are overlaid in gray. The dynamic folded energy-resolved pulse profile is shown in panel (g) with an energy resolution of 0.05\,keV. The relative amplitude of the pulse profiles as a function of energy is plotted in panel (h), which shows both the source properties and the detector response.}
	\label{Figure:Figure1}
\end{figure}

We folded all of the \textit{NICER} barycentric photon arrival times from XTE~J1810--197 using an ephemeris derived from contemporaneous radio pulsar timing measurements performed at 8.3\,GHz between 2019 February~7 (MJD\,58521) and 2019 February~26 (MJD\,58540)~(see Section~\ref{Section:PulsarTiming}; Methods). Relative phase shifts (in phase units) between the folded X-ray pulse profiles in the 1--2, 2--3, 3--4, 4--5, and 5--10\,keV energy bands were measured based on sinusoid fits to the pulse profiles in Figure~\ref{Figure:Figure1} (see Table~\ref{Table:Table3}). Our results indicate that the soft X-ray emission is nearly aligned between 1 and 10\,keV. This is consistent with the concentric geometry observed during the magnetar's 2003~outburst\citemain{Gotthelf+2005}, where the magnetar's thermal hot spot was surrounded by a larger, warmer emitting region. We measure a relative phase shift of --0.031\,$\pm$\,0.003 between the 3--5 and 5--10\,keV pulse profiles, which is notably smaller in magnitude than the $\Delta\phi$\,$\approx$\,0.1 phase shift reported in ref.\,\citemain{Gotthelf+2019} from a \textit{Nuclear Spectroscopic Telescope Array (NuSTAR)} observation on 2018 December 13 (MJD\,58465). If the apparent misalignment between the X-ray pulse profiles in ref.\,\citemain{Gotthelf+2019} was due to non-coaxial emission components, then our results indicate that the geometry of the X-ray emission has returned to being nearly concentric.

The dynamic energy-resolved folded light curve in Figure~\ref{Figure:Figure1}g shows that most of the X-ray photons are detected over a narrow energy range (between 1 and 4\,keV). Within this band, a significant fraction of the X-ray photons are detected with energies between roughly 1 and 2.5\,keV, where the XTI is most sensitive. The background-subtracted root-mean-squared~(RMS) X-ray pulsed fractions in the 0.5--5, 1--2, 2--3, 3--4, 4--5, and 5--10\,keV energy bands are listed in Table~\ref{Table:Table4}. These measurements show that the magnetar's pulsed fraction is increasing linearly as a function of energy between 1 and 5\,keV, which is consistent with an earlier \textit{NuSTAR} observation on 2018 December~13\citemain{Gotthelf+2019}.

\begin{figure}
	\centering
	\includegraphics[trim=0cm 0cm 0cm 0cm, clip=false, scale=0.4, angle=0]{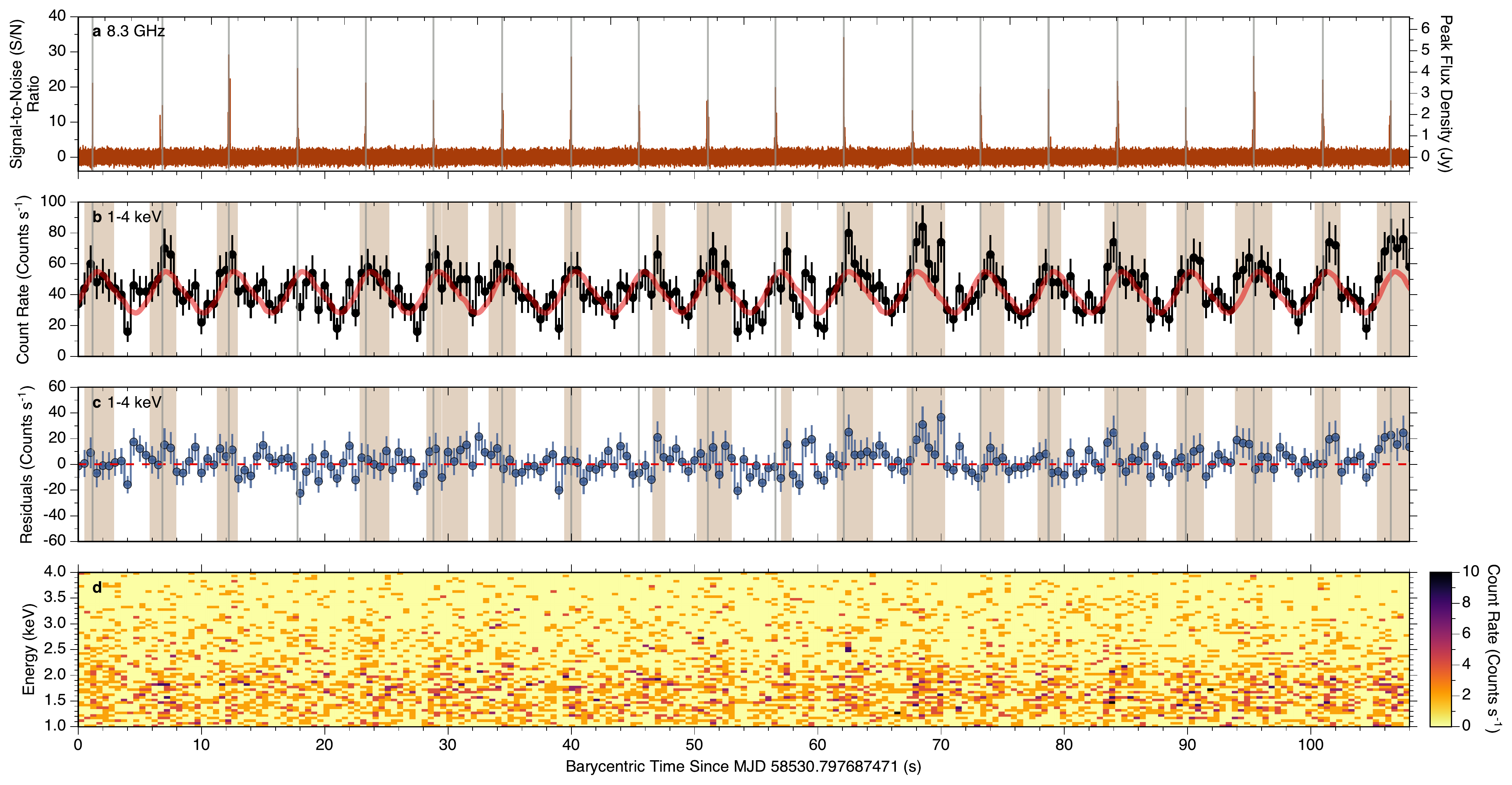}
	\caption{Simultaneous radio and X-ray pulses detected from XTE~J1810--197 on MJD\,58530. In panel (a), we show a series of (brown) radio pulses from DSN observations of the magnetar at 8.3\,GHz using 512\,$\mu$s time bins, along with (black) 1--4\,keV X-ray pulses simultaneously acquired with \textit{NICER} in panel (b) using 0.5\,s time bins. The smoothed, average X-ray pulse profile is overlaid in red in panel (b), after normalizing the pulse profile so that the area under the \textit{NICER} time-series and the smoothed profile are equal. The vertical gray lines in panels (a)--(c) indicate the peak time of each radio pulse during each rotation. The beige shaded regions in panels (b) and (c) denote the X-ray pulses identified by the zero-crossing algorithm described in Section~\ref{Section:ZeroCrossingAlgorithm}. The left edge, right edge, and width of the shaded regions correspond to the rising time, falling time, and duration of each X-ray pulse, respectively, as determined by the algorithm. We show the residuals, obtained by subtracting the (red) smoothed X-ray pulse profile from the (black) \textit{NICER} time-series, in blue in panel (c). The dynamic spectrum of the X-ray pulses is shown in panel (d) with an energy resolution of 0.05\,keV.}
	\label{Figure:Figure2}
\end{figure}

\begin{figure}
	\centering
	\includegraphics[trim=0cm 0cm 0cm 0cm, clip=false, scale=0.7, angle=0]{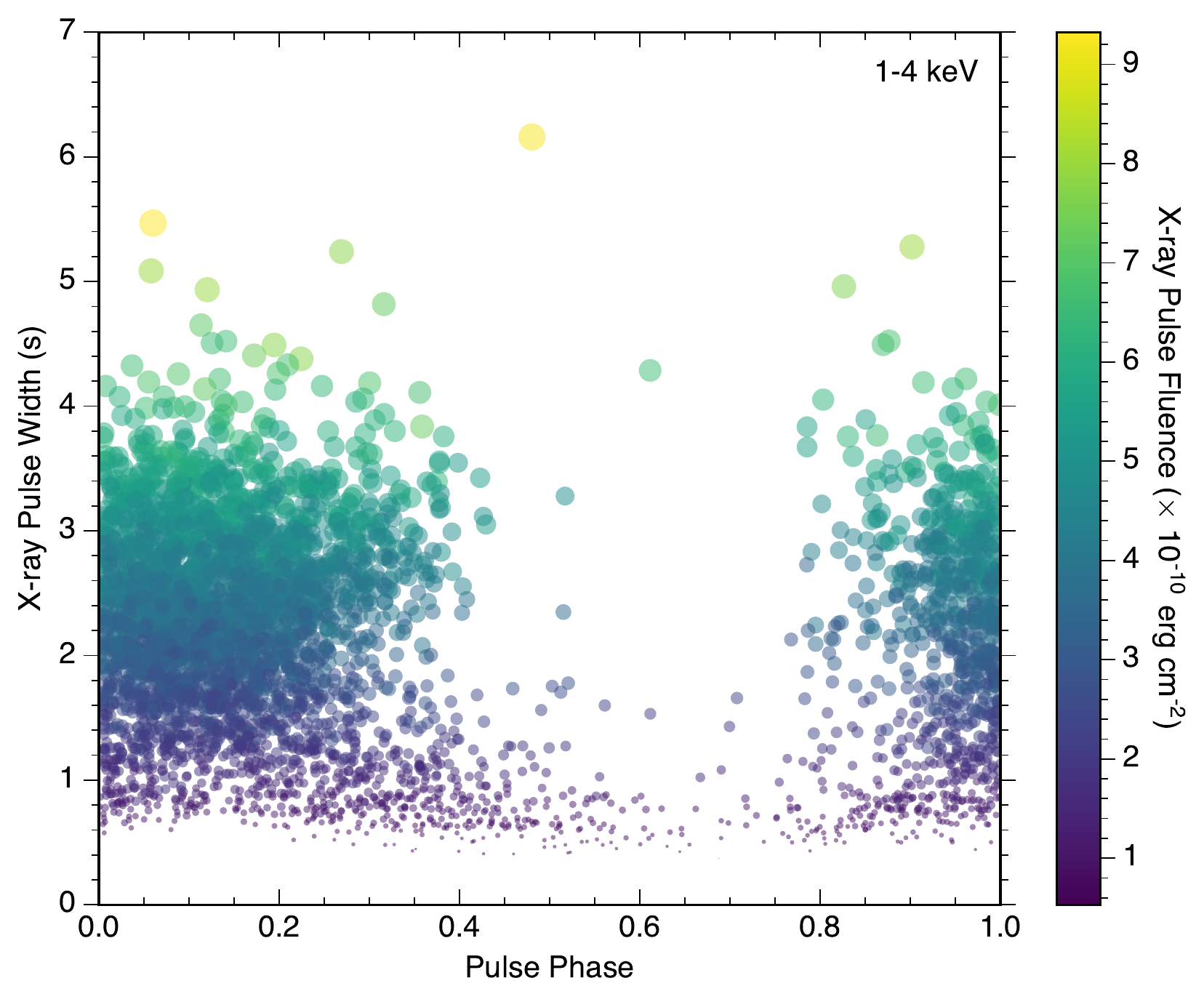}
	\caption{Temporal widths and fluences of the X-ray pulses detected by the zero-crossing algorithm as a function of XTE~J1810--197's rotational phase. These measurements were derived using \textit{NICER} observations between MJDs\,58520 and 58540 in the 1--4\,keV energy band. The color and size of each data point both represent the fluence of each X-ray pulse.}
	\label{Figure:Figure3}
\end{figure}

Bright X-ray pulses were detected from XTE~J1810--197 with \textit{NICER} during almost every rotation of the magnetar (e.g., see Figure~\ref{Figure:Figure2}). The X-ray pulses were identified by searching for temporal structure in the X-ray light curve using a zero-crossing algorithm. A description of the algorithm is provided in Section~\ref{Section:ZeroCrossingAlgorithm} (Methods). The X-ray pulses displayed statistically significant temporal variability on timescales shorter than the magnetar's rotational period, which is not due to Poisson fluctuations (see Section~\ref{Section:MonteCarloAnalysis}; Methods). A single bright X-ray pulse component was detected during most rotations, with a temporal width that varied between pulse cycles. However, X-ray pulses with multiple emission components were detected during $\sim$20\% of the rotational cycles, and approximately 20\% of the X-ray pulse components had temporal widths that were smaller than 1\,s (e.g., see Figure~\ref{Figure:Figure3}). The magnetar's individual X-ray pulses also showed pulse-to-pulse energy structure that was stochastically variable in time (e.g., see Figure~\ref{Figure:Figure2}d). Similar behavior was observed on 2019 February~25 during separate simultaneous X-ray and radio observations of the magnetar.

The X-ray pulses with larger widths originate from spin-modulated thermal emission from the surface of the magnetar, as the hot spots are swept across the line of sight. The presence of narrow width X-ray pulses indicates that there is also impulsive X-ray emission, which can be produced by external heating from relativistic magnetospheric particles bombarding the stellar surface\citemain{Beloborodov+2016, Ozel+2013}. This behavior indicates that the X-ray pulses are produced by quasi-thermal emission from one or multiple hot spots on the stellar surface\citemain{Gotthelf+2019, Guver+2019}. The X-ray pulses are therefore different from the impulsive, millisecond-wide pulses generated in the radio band (e.g., see Figure~\ref{Figure:Figure8}; Methods). These observations show that the thermal hot spots of magnetars can generate individually detectable X-ray pulses, which do not need to be induced by a giant flare.

We find that there is evidence of two different populations of X-ray pulses. In Figure~\ref{Figure:Figure3}, we show that the X-ray pulses with larger widths have higher fluences and are emitted over $\sim$60\% of the rotational phase range, while the narrow width X-ray pulses have lower fluences and are detected at virtually all rotational phases. The distribution of X-ray pulse fluences in Figure~\ref{Figure:Figure3} reveals a distinct separation between these two groups of X-ray pulses. We ascribe this behavior to anisotropic emission from the thermal regions, which can increase or reduce the apparent luminosity depending on the opacity in the magnetosphere and the inclination between the hot spots and the line of sight\citemain{Perna+2008}.

In Figure~\ref{Figure:Figure2}, we show a series of consecutive X-ray and radio pulses from a simultaneous observation with \textit{NICER} and the DSN on 2019 February~16. Although the peak times of the X-ray and radio pulses were nearly aligned during most rotational cycles, the X-ray/radio alignment was variable between subsequent rotations. Some pulse cycles revealed that the radio peak fell slightly before the X-ray peak, while other rotations showed that the radio peak time coincided with the X-ray peak or occurred shortly after. Approximately 65\% of the rotations shown in Figure~\ref{Figure:Figure2} had radio and X-ray peak times that agreed to within 0.5\,s (the time-resolution of the \textit{NICER} light curve).

\begin{figure}
	\centering
	\includegraphics[trim=0cm 0cm 0cm 0cm, clip=false, scale=0.76, angle=0]{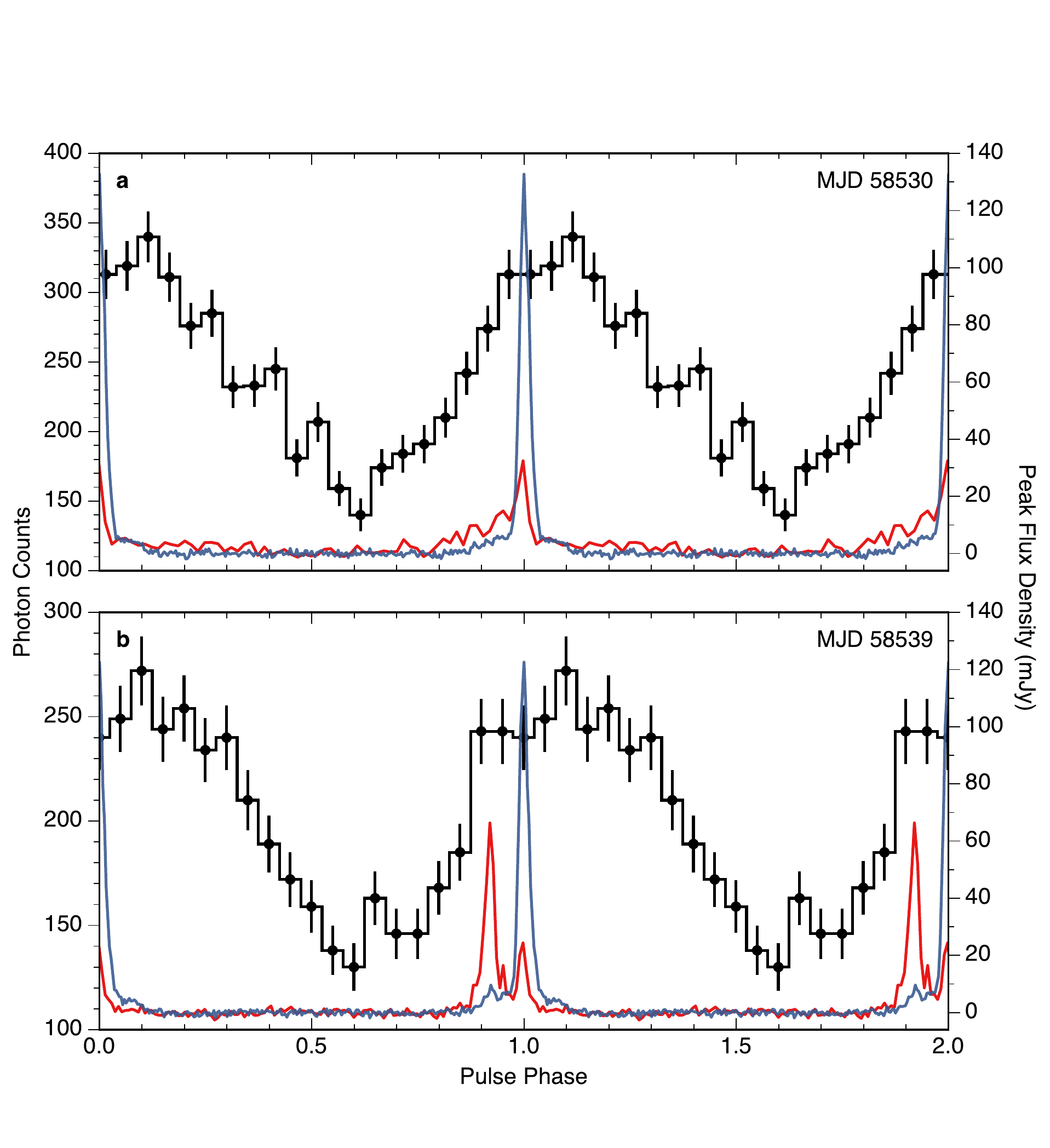}
	\caption{Folded X-ray and radio pulse profiles derived from simultaneous X-ray and radio observations of XTE~J1810--197 on MJDs~(a)~58530 and (b)~58539. The blue and red curves correspond to the average 8.3 and 31.9\,GHz radio pulse profiles of the magnetar, respectively. The black curves show the \textit{NICER} 1--4\,keV pulse profiles, folded with 20 phase bins using a phase-connected radio ephemeris spanning each X-ray observation. Phase~0 in panels (a) and (b) correspond to MJDs\,58530.761334907~(TDB) and 58539.774091226~(TDB), respectively.}
	\label{Figure:Figure4}
\end{figure}

The folded 1--4\,keV X-ray pulse profiles from 2019 February~16 and 2019 February~25 are shown in Figure~\ref{Figure:Figure4}, along with the average pulse profiles from simultaneous radio observations at 8.3 and 31.9\,GHz. In order to align the X-ray and radio pulse profiles on each day, we folded the radio and X-ray data using measurements of the magnetar's rotational period during each individual observation, which were derived from a phase-coherent timing solution using pulse times of arrival (ToAs) at 8.3\,GHz (see Table~\ref{Table:Table2}). On 2019 February~16 and 2019 February~25, the 8.3\,GHz pulse profile peak was offset from the peak of the X-ray pulse profile by $\Delta\phi$\,$=$\,0.11\,$\pm$\,0.05 and $\Delta\phi$\,$=$\,0.10\,$\pm$\,0.05, respectively. Although these phase shifts are comparable to the X-ray/radio phase alignment reported during XTE~J1810--197's 2003 and 2018~outbursts ($\Delta\phi_{\text{2006}}$\,$=$\,0.167\,$\pm$\,0.006\,\citemain{Camilo+2007a} and $\Delta\phi_{\text{2018}}$\,$\approx$\,0.13\,\citemain{Gotthelf+2019}), we note that these values correspond to the average alignment, which can differ from the alignment during individual rotations. A strong precursor component was also present in the 31.9\,GHz pulse profile on 2019 February~25, which was not seen in the profile $\sim$9 days earlier. The phase offset between the 31.9\,GHz precursor component and the peak of the 8.3\,GHz pulse profile is $\Delta\phi$\,$=$\,0.080\,$\pm$\,0.009. The structure preceding the 31.9\,GHz average pulse profile peak on 2019 February~16 is attributed to a population of radio pulses with lower flux densities (see Figure~\ref{Figure:Figure7}c; Methods).

Between 2019 February~6 and 2019 February~26 (MJDs 58520--58540), XTE~J1810--197's 1--4\,keV absorbed X-ray flux decayed from 1.45\,$\times$\,10$^{\text{--10}}$ to 1.13\,$\times$\,10$^{\text{--10}}$\,erg\,s$^{\text{--1}}$\,cm$^{\text{--2}}$ at an average rate of (--1.21\,$\pm$\,0.04) $\times$\,10$^{\text{--12}}$\,erg\,s$^{\text{--1}}$\,cm$^{\text{--2}}$\,day$^{\text{--1}}$. During this time period, the peak absorbed X-ray fluxes of the detected X-ray pulses ranged between 1.3\,$\times$\,10$^{\text{--10}}$ and 2.8\,$\times$\,10$^{\text{--10}}$\,erg\,s$^{\text{--1}}$\,cm$^{\text{--2}}$. Assuming that the X-ray emission was produced by a blackbody emitting region at a distance of 3.5\,kpc, with an area of $A_{\text{BB}}$\,$=$\,213\,km$^{\text{2}}$ (see Section~\ref{Section:XRayFluxEnergy}; Methods), we find that the peak X-ray luminosities of the X-ray pulses (averaged over the emitting area; $\langle L_{\text{X}}\rangle$\,$=$\,$L_{\text{X}}$\,$\left(\frac{A_{\text{BB}}}{\text{1\,km}^{\text{2}}}\right)^{\text{--1}}$) were between 0.9\,$\times$\,10$^{\text{33}}$ and 1.9\,$\times$\,10$^{\text{33}}$\,erg\,s$^{\text{--1}}$\,km$^{\text{--2}}$. From the peak X-ray luminosities of the X-ray pulses, we find that the average effective surface temperature over the emitting region ($T_{\text{s}}$\,$=$\,$(\langle L_{\text{X}}\rangle$\,$/$\,$\sigma_{\text{SB}})^{1/4}$) is approximately \\ $T_{\text{s}}$\,$\approx$\,(6--8)\,$\times$\,10$^{\text{6}}$\,K, where $\sigma_{\text{SB}}$\,$=$\,5.67\,$\times$\,10$^{\text{--5}}$\,erg\,s$^{\text{--1}}$\,cm$^{\text{--2}}$\,K$^{\text{--4}}$ is the Stefan-Boltzmann constant. The inferred total radiative energies of the X-ray pulses, averaged over the emitting surface ($\langle E_{\text{total}}\rangle$ $=$\,$E_{\text{total}}$\,$\left(\frac{A_{\text{BB}}}{\text{1\,km}^{\text{2}}}\right)^{\text{--1}}$), were (0.4--6.4)\,$\times$\,10$^{\text{33}}$ erg\,km$^{\text{--2}}$.

The energetics, widths, and morphology of XTE~J1810--197's X-ray pulses are remarkably different from those previously observed from giant flares and short X-ray bursts from magnetars. Giant flares from magnetars (e.g., see refs.\,\citemain{Hurley+1999, Palmer+2005}) are typically characterized by an initial spike lasting $\sim$10--100\,ms, followed by an exponential decaying tail lasting several minutes, with peak X-ray luminosities in the range of 10$^{\text{44}}$--10$^{\text{47}}$\,erg\,s$^{\text{--1}}$. These events are rare and occur roughly once per decade. Short X-ray bursts from magnetars (e.g., see ref.\,\citemain{Woods+2005}) have burst durations that range from a few milliseconds to a few seconds, with tails that can sometimes last several minutes. The peak X-ray luminosities of short X-ray bursts can range between 10$^{\text{36}}$ and 10$^{\text{43}}$\,erg\,s$^{\text{--1}}$. Therefore, the X-ray pulses reported here from XTE~J1810--197 are temporally distinct and less energetic than giant flares and X-ray bursts previously observed from magnetars. Moreover, we find that the pulse-energy distributions of the X-ray and radio pulses are characterized by different statistical distributions (see Section~\ref{Section:PulseEnergyDistributions} and Figure~\ref{Figure:Figure5}).

The persistent emission of X-ray pulses during \textit{NICER} observations spanning $\sim$20\,days, along with the derived surface temperatures from the luminosities of the X-ray pulses, indicate that the thermal regions producing the emission are heated quasi-steadily. This behavior can be explained by external heating and is consistent with predictions from the twisted magnetosphere model used in the past to explain XTE~J1810--197's radiative behavior during its 2003~outburst\citemain{Beloborodov2009, Beloborodov+2016}. In this model, strong twists and powerful currents in the magnetosphere are generated by the evolution and decay of an ultra-strong magnetic field anchored in the magnetar's crust. The untwisting process forms a current-carrying bundle of field lines, known as the $j$-bundle, which powers the magnetar's emission on the untwisting timescale of months to years\citemain{Beloborodov+2016}. A hot spot is created at the footprint of the $j$-bundle as the stellar surface is heated by the bombardment of relativistic magnetospheric particles, and a significant fraction of the dissipated power can be radiated quasi-thermally\citemain{Beloborodov+2016}. The stellar surface is also expected to be thermally heated via anisotropic heat conduction through the neutron star's crust due to the presence of strong sub-surface magnetic fields.

The alignment between the individual X-ray and 8.3\,GHz radio pulses suggests that they both originate near the same portion of the neutron star. We attribute the variability in the pulse-to-pulse alignment of the X-ray/radio pulses and the changes in the temporal structure of the X-ray pulses to fluctuations in the thermal emission from the magnetar's hot spots. Particle bombardment from returning magnetospheric currents can externally heat the hot spots on the neutron star's surface on sub-rotational timescales (e.g., see ref.\,\citemain{Ozel+2013}).

We did not find evidence of a correlation between the temporal structure or peak amplitudes of the X-ray and radio pulses (e.g.,~see Figure~\ref{Figure:Figure2}). This indicates that the magnetar's radio emission is uncorrelated with its persistent soft X-ray emission on rotational timescales. Previously, simultaneous suppression of radio emission was reported during short magnetar-like X-ray bursts from PSR~J1119--6127, a high magnetic field radio pulsar\citemain{Archibald+2017}. This was attributed to the ejection of a pair-plasma fireball into the magnetosphere, which is thought to quench the radio emission by shielding the electric field in the particle accelerating region and then recover on timescales of 10--100\,s\citemain{Yamasaki+2019}. However, similar behavior was not detected during our simultaneous X-ray and radio observations of XTE~J1810--197.

The radio pulses from XTE~J1810--197 share similarities with some of the radio bursts previously detected from repeating FRB sources. In particular, the magnetar's 8.3\,GHz radio pulses display frequency structure that is not observed in the radio pulses detected simultaneously at 31.9\,GHz. Additionally, some of the magnetar's radio pulses were not simultaneously detected at both radio frequencies~(e.g., see Figure~\ref{Figure:Figure8}). This indicates that many of XTE~J1810--197's radio pulses are not broadband and have a spectral index that varies between pulse components. Similar behavior has been observed in bursts from repeating FRBs, such as FRB~121102. In Section~\ref{Section:FastRadioBursts} (Methods), we further describe the morphology of XTE~J1810--197's radio pulses and discuss possible links with the emission from repeating FRB sources. Although the luminosities of XTE~J1810--197's radio pulses are inconsistent with the energy output of bursts from repeating FRBs, such as those from FRB~121102 and FRB~180916.J0158+65, we note that a $\gtrsim$\,1.5\,MJy\,ms radio burst was recently detected from the active magnetar SGR~1935+2154\citemain{Bochenek+2020b}. This suggests that active magnetars are able to produce sufficiently energetic radio bursts that may explain some extragalactic FRBs. In addition, an X-ray burst was also detected contemporaneously with this high fluence radio burst from SGR~1935+2154 (e.g., see ref.\,\citemain{Mereghetti+2020b}). Therefore, our simultaneous observations of individual radio and X-ray pulses from XTE~J1810--197 during its recent outburst are important for characterizing the behavior of active magnetars, which are now thought to be a source of some extragalactic FRBs.

\newpage

\newgeometry{top=1cm}

\newgeometry{top=2.5cm}

\begin{addendum}

\item[Acknowledgments] The authors would like to thank Ben Stappers for providing a radio ephemeris of the Crab pulsar and Jim Palfreyman for supplying a radio ephemeris of the Vela pulsar. We are also grateful to Wenbin Lu for valuable discussions. \\

A.B.P. acknowledges support by the Department of Defense (DoD) through the National Defense Science and Engineering Graduate (NDSEG) Fellowship Program and by the National Science Foundation (NSF) Graduate Research Fellowship under Grant No. DGE-1144469. T.E. is supported by JSPS/MEXT~KAKENHI grant number 18H01246. \\
 
We thank the Jet Propulsion Laboratory's Research and Technology Development program and Caltech's President's and Director's Fund for partial support at JPL and the Caltech campus. A portion of this research was performed at the Jet Propulsion Laboratory, California Institute of Technology and the Caltech campus under a Research and Technology Development Grant through a contract with the National Aeronautics and Space Administration (NASA). U.S. government sponsorship is acknowledged. We acknowledge support from the DSN team for scheduling and carrying out the radio observations. We also thank Charles Lawrence for providing programmatic support for this work. \\

This work was supported by NASA through the \textit{NICER} mission and the Astrophysics Explorers Program, and made use of data and software provided by the High Energy Astrophysics Science Archive Research Center (HEASARC). Portions of this work performed at the United States Naval Research Laboratory (NRL) were supported by NASA.

\item[Competing Interests] The authors declare that they have no
competing financial interests.

\item[Contributions] A.B.P. was the primary author of the manuscript, led the X-ray and radio data analysis, and produced the figures. A.B.P. carried out the pulsar timing and pulsation analyses with contributions from W.A.M. A.B.P., P.S.R., W.A.M., and J.K. aligned the folded X-ray and radio light curves. W.A.M. and C.J.N. coordinated the radio and X-ray observations with S.H., Z.A., and K.C.G. A.B.P., W.A.M., T.A.P., P.S.R., T.E., T.G., Z.A., and W.C.G.H. contributed to the physical interpretation of the results. A.B.P., W.A.M., T.A.P., P.S.R., J.K., S.H., T.E., T.G., Z.A., and W.C.G.H. edited the manuscript. W.A.M. is the PI of a DSN proposal focused on monitoring XTE~J1810--197 during its recent outburst. W.A.M., T.E., and T.G. are co-PIs of the \textit{NICER} Target of Opportunity (ToO) proposal. K.C.G. is the PI of \textit{NICER}, Z.A. is the Deputy PI and Science Lead for \textit{NICER}, and T.E. is the leader of the magnetar and magnetosphere science subgroup of \textit{NICER}. T.A.P. is A.B.P.'s Ph.D. advisor.

\item[Correspondence] Correspondence and requests for materials should be addressed to A.B.P.\\(email: aaron.b.pearlman@caltech.edu).

\clearpage

\end{addendum}

\bibliographystylemain{naturemag}
\bibliographymain{references}

\clearpage

\begin{methods}

\section{X-ray Observations}
\label{Section:XrayObservations}

\indent\indent We observed XTE~J1810--197 with the \textit{NICER} X-ray Timing Instrument~(XTI)\citemethods{Gendreau+2016} on board the \textit{International Space Station} between 2019 February~6 (MJD\,58520) and 2019 February~26 (MJD\,58540), soon after the magnetar's position was sufficiently offset from the Sun. The XTI consists of an aligned array of 56 X-ray ``concentrator'' optics and silicon drift detectors (52 operational on orbit), which is sensitive to soft X-ray photons between 0.2 and 12\,keV and has a large effective area of $\sim$1900\,cm$^{\text{2}}$ at 1.5\,keV. The precision timing capabilities of the XTI enable the arrival times of individual X-ray photons to be measured to an accuracy better than 100\,ns. The X-ray data were processed using the \textit{NICER} data analysis software\customfootnote{1h}{~See https://heasarc.gsfc.nasa.gov/docs/nicer/nicer\_analysis.html.} (DAS version 2018-11-19 V005a). We cleaned the data using the standard \textit{NICER} calibration and filtered out times with a high background count rate using the niprefilter2 and nimaketime routines. We excluded event times near the South Atlantic Anomaly~(SAA), when the angular pointing separation was larger than 0.015$^{\circ}$, and when the elevation angle was less than 30$^{\circ}$ above the limb of the Earth or less than 40$^{\circ}$ above the bright Earth limb. We also removed ``hot'' detectors from our analysis, which were flagged when an individual detector recorded more events than 3$\sigma$ above the mean number of events across all of the detectors. The event times were then corrected to the solar system barycenter using the barycorr FTOOLS\citemethods{Blackburn1995}\textsuperscript{,}\customfootnote{2h}{~See http://heasarc.gsfc.nasa.gov/ftools.} routine and the Jet Propulsion Laboratory~(JPL) DE-405 ephemeris.

Since \textit{NICER} is a non-imaging X-ray telescope, the background count rate was estimated using a space weather-based spectral background model derived from observations of ``blank sky'' fields\citemain{Guver+2019}. A background X-ray spectrum, which incorporated contributions from the time-dependent particle background, optical loading from the Sun, and the diffuse sky background, was used to determine the energy-dependent background count rates. The calculated background count rates in the 0.5-5, 1--2, 2--3, 3--4, 4--5, and 5--10\,keV energy bands were 0.486, 0.153, 0.076, 0.054, 0.045, and 0.169 counts\,s$^{\text{--1}}$, respectively. Between 0.5 and 5\,keV, the ratio between the background and source count rates was $\lesssim$\,1\%. Since the background is negligible in this energy range, we did not perform further background subtraction. However, the X-ray pulsed fractions of XTE~J1810--197, provided in Table~\ref{Table:Table4}, have been corrected for the energy-dependent background count rate.

\section{Radio Observations}
\label{Section:RadioObservations}

\indent\indent High frequency radio observations of XTE~J1810--197 were carried out using the NASA~DSN 34\,m radio telescopes (DSS-34 and DSS-35)\citemain{Pearlman+2019} near Canberra, Australia on 2019~February~16 (MJD\,58530) and 2019~February~25 (MJD\,58539). On both days, dual circular polarization data were simultaneously recorded at central radio frequencies of 8.3 and 31.9\,GHz, with roughly 350\,MHz of bandwidth. Power spectral density measurements across the bands were channelized and saved in a digital polyphase filterbank with a time and frequency resolution of 512\,$\mu$s and $\sim$1\,MHz, respectively. The elevation-corrected system temperatures ($T_{\text{sys}}$) at 8.3/31.9\,GHz during the radio observations on MJDs\,58530 and 58539 were 24(5)/46(9)\,K and 22(4)/41(8)\,K, respectively. The uncertainties correspond to 20\% errors on the system temperature values.

The data were cleaned by first removing spurious signals due to radio frequency interference~(RFI) using the rfifind tool from the PRESTO pulsar search software package\citemethods{Ransom2001}\textsuperscript{,}\customfootnote{3h}{~See https://www.cv.nrao.edu/$\sim$sransom/presto.}. Next, the bandpass response was flattened, and then we subtracted the moving average from each data point using 10\,s of data around each sample to remove low frequency variations from the baseline of each frequency channel. We used the TEMPO\customfootnote{4h}{~See http://tempo.sourceforge.net.} timing analysis software package to correct the sample times to the solar system barycenter and then incoherently dedispersed the data using the magnetar's nominal dispersion measure~(DM) of 178\,pc\,cm$^{\text{--3}}$.

\section{Pulsar Timing}
\label{Section:PulsarTiming}

\indent\indent Phase-connected timing solutions spanning our radio observations on 2019 February~16 and 2019 February~25 were obtained using 8.3\,GHz observations of XTE~J1810--197. We derived ToAs by cross-correlating individual measured profiles, which were constructed by folding sub-integrations of the magnetar, in the Fourier frequency domain\citemethods{Taylor1992} using a standard template based on the average pulse profile. The ToAs were calculated using the get\_TOAs.py tool from PRESTO\citemethods{Ransom2001} and fit using the TEMPO2\customfootnote{5h}{~See https://www.atnf.csiro.au/research/pulsar/tempo2.} timing analysis software package\citemethods{Hobbs+2006}. Separate timing solutions were derived for each of our radio observations by fitting only for the magnetar's spin frequency, $\nu$. Additional spin frequency derivatives were not needed to obtain a phase-connected timing solution during each epoch. We fixed the position of the magnetar to the value reported in ref.\,\citemethods{Helfand+2007} from Very Long Baseline Array~(VLBA) observations. Due to the variability in the magnetar's pulse intensity at 31.9\,GHz, we were unable to carry out a multi-frequency timing analysis. Since the timing models were based on ToAs from a single observing frequency, we set the pulsar DM to be equal to the magnetar's nominal DM in these models. Any alignment error between the 8.3 and 31.9\,GHz radio pulse profiles due to uncertainty in the DM is negligible. We also increased the uncertainties on the ToAs by multiplying the error on each ToA by a scaling error factor~(EFAC) value, given by $\epsilon$\,$=$\,$\sqrt{\chi_{\nu}^{2}}$, which yielded a reduced chi-squared value of $\chi_{\nu}^{2}$\,$=$\,1 by construction in all of our models. This is a standard technique used in pulsar timing to ensure more realistic parameter uncertainties since the errors on ToAs obtained from cross-correlating profiles with templates are often underestimated\citemethods{Scholz+2017}. The timing solutions obtained using this procedure are provided in Table~\ref{Table:Table2}. All reference times are barycentric and scaled to infinite frequency.

These timing solutions were used to derive the phase alignment between the folded X-ray and radio pulse profiles shown in Figure~\ref{Figure:Figure4}. The accuracy of XTE~J1810--197's X-ray/radio phase alignment was verified using \textit{NICER} and DSN observations of the Crab pulsar, carried out close in time. We confirmed that the peaks in the Crab pulsar's X-ray and radio pulse profiles were phase aligned, which is consistent with previous measurements of the pulsar's X-ray/radio phase alignment (e.g., see ref.~\citemethods{Abdo+2010}). Radio observations, taken close in time, of the Vela pulsar were also carried out using the DSN and the Mount Pleasant 26\,m radio telescope, located near Hobart, Tasmania. We found that radio timing measurements of the Vela pulsar from both observatories were in agreement. Both of these independent tests indicate that there are not any significant systematic errors in our absolute timing accuracy that would affect the alignment shown in Figure~\ref{Figure:Figure4}.

The X-ray data in Figures~\ref{Figure:Figure1} and~\ref{Figure:Figure7}a were folded using an ephemeris derived from spin frequency measurements of XTE~J1810--197 on 2019 February~7 (MJD\,58521), 2019 February~16 (MJD\,58530), and 2019 February~25 (MJD\,58539). The rotational frequency on each day was obtained from 8.3\,GHz observations of the magnetar using the pulsar timing procedure described above. The ephemeris was constructed by fitting a linear polynomial to these measurements using a nonlinear least-squares fitting procedure. The parameters comprising the ephemeris are $\nu_{0}$\,$=$\,0.18045678(4)\,Hz and $\dot{\nu}$\,$=$\,(--4.8\,$\pm$\,0.7) \\ $\times$\,10$^{\text{--13}}$\,Hz\,s$^{\text{--1}}$, where $\nu_{0}$ is the rotational frequency of the magnetar on MJD\,58530.761334907 (TDB) and $\dot{\nu}$ is the time derivative of the magnetar's spin frequency. We note that our measurement of the magnetar's spin frequency derivative is roughly twice as large in magnitude as the value reported in ref.\,\citemethods{Levin+2019}, which is likely explained by variability in the magnetar's spin-down torque. Rapid changes in the neutron star torque have also been observed from other magnetars following an outburst (e.g., refs.\,\citemain{Ibrahim+2004, Camilo+2007a}\textsuperscript{,}\citemethods{Dib+2012, Camilo+2018, Enoto+2019}). \\ \\

\section{X-ray Pulse Analysis}
\label{Section:XRayAnalysis}

\subsection{Zero-Crossing Algorithm}
\label{Section:ZeroCrossingAlgorithm}

Since the X-ray pulse structure of XTE~J1810--197 was variable between rotational cycles, we used a zero-crossing algorithm to identify individual pulsed emission components containing X-ray temporal structure. We searched for X-ray pulses using the 1--4\,keV \textit{NICER} light curve in order to maximize the signal-to-noise~(S/N) ratio of each pulse. The cleaned event light curve was first binned to a time resolution of 0.5\,s and divided into smaller light curve segments, which each consisted of times when \textit{NICER} was continuously pointed at the magnetar. After subtracting the mean count rate from each of the light curve segments, we searched for times when the product of the count rates in adjacent time bins was negative. This indicated that the relative count rate had changed from a negative to positive value or from a positive to negative value. The slope of the light curve during these time intervals allowed us to determine whether these intervals contained a rising or falling time. The precise rising and falling times were obtained by linearly interpolating between time bins where the relative count rate in the light curve changed sign. The complex variability in Figure~\ref{Figure:Figure2}b shows that this technique is superior to searching for pulses by imposing a minimum or maximum count rate threshold. The times when the mean-subtracted signal changes sign are well-behaved, while imposing a threshold in the search algorithm would cause some events to be excluded.

In our analysis, we define the peak amplitude of each X-ray pulse to be the maximum count rate above the mean level during an adjacent set of rising and falling times, and the arrival time of each X-ray pulse corresponds to the time when this maximum occurred. These measurements are displayed in Figure~\ref{Figure:Figure7}a. The estimated pulse width of each event is given by the difference between the rising and falling times surrounding the X-ray pulse. The fluence or energy of each pulse is calculated from the time-integrated X-ray flux of each event. The pulse-energy distribution of the X-ray pulses is shown in Figure~\ref{Figure:Figure5}c.

\subsection{Monte Carlo Analysis}
\label{Section:MonteCarloAnalysis}

The X-ray pulses from XTE~J1810--197 displayed significant temporal variability between subsequent rotations of the neutron star (e.g., see Figure~\ref{Figure:Figure2}b). To demonstrate that the changes in the X-ray pulse structure were not dominated by Poisson fluctuations, we performed a Monte Carlo analysis where we generated 10$^{\text{6}}$ simulated X-ray pulses for each of the 4,013 X-ray pulses detected by the zero-crossing algorithm. Each simulated X-ray pulse was derived by Poisson sampling the average 1--4\,keV X-ray pulse profile after scaling the profile such that its amplitude was equal to the peak amplitude of an observed X-ray pulse and interpolating the profile to a time resolution of 0.5\,s. The number of counts in each time bin of the simulated X-ray pulses was determined by randomly sampling from a Poisson distribution parameterized by the number of counts in the corresponding time bin of the scaled and interpolated average 1-4\,keV X-ray pulse profile. The simulated X-ray pulses were then analyzed by the zero-crossing algorithm, and we recorded the rising time, falling time, peak time, and peak amplitude of each significantly detected X-ray pulse component.

A single X-ray pulse component was significantly detected in $>$\,99.9\% of the simulated X-ray pulses analyzed by the zero-crossing algorithm. This demonstrates that the zero-crossing algorithm reliably detects X-ray pulse components with high fidelity. The pulse width of each simulated X-ray pulse component was calculated from the difference between the rising and falling times. In Figure~\ref{Figure:Figure6}, we show the observed and simulated X-ray pulse width distributions. The simulated X-ray pulse width distribution is considerably narrower than the observed distribution and peaked at approximately half the magnetar's rotational period. The simulated distribution also shows a small number (0.08\%) of pulse components with widths less than 1.5\,s, which are attributed to false-positive detections of low amplitude pulse components due to Poisson variability. The observed X-ray pulse width distribution is quasi-bimodal and much broader than the simulated distribution. The notable differences between the observed and simulated X-ray pulse width distributions also demonstrate that there are statistically significant deviations between the temporal structure observed in the individual X-ray pulses and the average X-ray pulse shape.

\subsection{X-ray Flux, Luminosity, and Total Energy}
\label{Section:XRayFluxEnergy}

In order to measure the X-ray fluxes of the X-ray pulses, a blackbody plus power-law model was fit to the background-subtracted spectra derived from each of the event light curves used in the zero-crossing analysis. The spectra were formed by first extracting photons in the 1--4\,keV energy band. Each spectrum was then binned so that there were at least 50 counts in each spectral channel after background subtraction. We fixed the hydrogen column density ($N_{\text{H}}$) in the fits to be 1.35\,$\times$\,10$^{\text{22}}$\,cm$^{\text{2}}$, the same value obtained from the blackbody plus power-law fits in ref.~\citemain{Guver+2019}. Conversion factors for each event light curve were determined from the model-predicted absorbed X-ray fluxes and count rates. The peak absorbed X-ray flux of each X-ray pulse was determined by multiplying the peak count rate by its corresponding conversion factor.

We obtained an average blackbody emitting area of 213\,$\pm$\,9\,km$^{\text{2}}$ between MJDs 58520 and 58540 from the spectral fits, which implies an average apparent blackbody emitting radius of 5.1\,$\pm$\,0.1\,km. The peak X-ray luminosities and total radiative energies of the X-ray pulses, averaged over the emitting surface, were determined using this blackbody emitting area, assuming a distance of 3.5\,kpc to the magnetar.

\section{Radio Single Pulse Analysis}
\label{Section:RadioSinglePulseAnalysis}

\subsection{Matched Filtering Algorithm}
\label{Section:MatchedFilteringAlgorithm}

We used a Fourier domain matched filtering algorithm\citemain{Pearlman+2018}\textsuperscript{,}\citemethods{Ransom2001} to search for single pulses in our simultaneous 8.3 and 31.9\,GHz radio observations of XTE~J1810--197. After masking bad data corrupted by~RFI and applying the bandpass and baseline corrections described in Section~\ref{Section:RadioObservations}, we barycentered and incoherently dedispersed the data using the magnetar's nominal DM. The radio pulses were then identified by convolving the full resolution time series data with boxcar functions with widths ranging from 512\,$\mu$s to 153.6\,ms. If a radio pulse from the same section of data was detected with multiple boxcar widths, we only recorded the highest S/N event in the final list. All events with S/N\,$\ge$\,7.0 were stored for further analysis. The peak flux density of each event was calculated using the radiometer equation\citemethods{McLaughlin+2003}:
\begin{equation}
S_{\text{peak}}=\frac{\beta\,T_{\text{sys}}(\text{S/N})_{\text{peak}}}{G\,\sigma_{\text{off}}\sqrt{\Delta\nu\,n_p\,t_{\text{peak}}}},
\label{Equation:PeakFluxDensity}
\end{equation}
where $\beta$\,$\approx$\,1 is a correction factor that accounts for system imperfections such as digitization of the signal, $T_{\text{sys}}$ is the effective system temperature, (S/N)$_{\text{peak}}$ is the peak S/N of the radio pulse, $G$\,$=$\,0.24\,K\,Jy$^{\text{--1}}$ is the telescope gain, $\sigma_{\text{off}}$\,$=$\,1 is the off-pulse standard deviation, $\Delta\nu$ is the observing bandwidth, $n_p$ is the number of polarizations, and $t_{\text{peak}}$ denotes the integration time at the peak of the radio pulse. The fluence of each event was determined from the time-integrated flux density.

\section{Radio and X-ray Pulse-Energy Distributions}
\label{Section:PulseEnergyDistributions}

\indent\indent Pulse-energy distributions of the radio pulses detected at 8.3 and 31.9\,GHz on MJD\,58530 are shown in Figures~\ref{Figure:Figure5}a and~\ref{Figure:Figure5}b, respectively. Scaled log-normal distributions, given by:
\begin{equation}
P_{\text{log-norm}}\left(x=\frac{E}{\langle E\rangle}\right)=\frac{C}{\sqrt{2\pi}\sigma_{\text{log-norm}}x}\exp\left[-\frac{(\ln x-\mu_{\text{log-norm}})^{2}}{2\sigma_{\text{log-norm}}^{2}}\right],
\label{Equation:ScaledLogNormalDist}
\end{equation}
where $x$ is the normalized energy of each radio pulse, $C$ is a scaling factor, $\mu_{\text{log-norm}}$ is the mean of the distribution, and $\sigma_{\text{log-norm}}$ is the standard deviation of the distribution, were fit to each of these distributions using a weighted nonlinear least-squares fitting procedure. The 8.3\,GHz pulse-energy distribution is well-described by a log-normal distribution \\ ($\chi^{2}_{\text{red}}$\,$=$\,0.98, dof\,$=$\,18) with $\mu_{\text{log-norm}}$\,$=$\,--0.218\,$\pm$\,0.009 and $\sigma_{\text{log-norm}}$\,$=$\,0.709\,$\pm$\,0.008. We obtained a best-fit log-normal distribution ($\chi^{2}_{\text{red}}$\,$=$\,0.36, dof\,$=$\,6) with $\mu_{\text{log-norm}}$\,$=$\,--0.09\,$\pm$\,0.05 and $\sigma_{\text{log-norm}}$\,$=$\,0.48\,$\pm$\,0.04 after fitting the 31.9\,GHz pulse-energy distribution. These best-fit log-normal distributions are overlaid in red in Figures~\ref{Figure:Figure5}a and~\ref{Figure:Figure5}b. We obtained a smaller $\chi^{2}_{\text{red}}$ value from the log-normal fit to the 31.9\,GHz pulse-energy distribution since fewer radio pulses were detected at this frequency, which resulted in large relative error bars. However, we note that interstellar scintillation~(ISS) and/or intrinsic variability (e.g., pulse nulling) may have affected the shape of the 31.9\,GHz pulse-energy distribution. For comparison, during the 2003~outburst, XTE~J1810--197's single pulse emission displayed both log-normal and power-law behavior between 1.4 and 8.35\,GHz\citemethods{Serylak+2009}.

The pulse-energy distribution of all of the X-ray pulse components, along with the distribution derived from selecting only the brightest component during each rotation, is shown in Figure~\ref{Figure:Figure5}c. We fit a scaled Gaussian distribution, given by:
\begin{equation}
P_{\text{gauss}}\left(x=\frac{E}{\langle E\rangle}\right)=\frac{C}{\sqrt{2\pi}\sigma_{\text{gauss}}}\exp\left[-\frac{(x-\mu_{\text{gauss}})^{2}}{2\sigma_{\text{gauss}}^{2}}\right],
\label{Equation:GaussianDist}
\end{equation}
to the pulse-energy distribution of the brightest pulse components with $x$\,$\gtrsim$\,0.6. The data are well-modeled by a Gaussian distribution ($\chi^{2}_{\text{red}}$\,$=$\,0.91, dof\,$=$\,11), with a mean and standard deviation of $\mu_{\text{gauss}}$\,$=$\,1.12\,$\pm$\,0.01 and $\sigma_{\text{gauss}}$\,$=$\,0.384\,$\pm$\,0.009, respectively. This indicates that the energetics of the brightest X-ray pulses from the magnetar's hot spot are well-described by a Gaussian process. These results also show that the magnetar's X-ray pulse-energy distribution differs from its radio pulse-energy distribution.

\section{Radio Pulse Morphology}
\label{Section:RadioPulseMorphology}

\indent\indent Our high frequency radio observations of XTE~J1810--197 reveal that the magnetar is emitting bright single pulses with multiple narrow emission components following its recent reactivation. This behavior is similar to the emission characteristics observed during the magnetar's 2003~outburst\citemain{Camilo+2006}. Although many of these pulse components were simultaneously detected at 8.3 and 31.9\,GHz, not all of the emission components were detected at both frequencies (e.g., see Figure~\ref{Figure:Figure8}). This suggests that either a substantial fraction of the magnetar's single pulse components have a steep and variable radio spectrum or they are not all emitted over a broadband frequency range.

We folded the ToAs of the single pulse emission components detected at 8.3 and 31.9\,GHz using the radio ephemeris described in Section~\ref{Section:PulsarTiming}. The time-phase distributions of these events are shown in Figures~\ref{Figure:Figure7}b and~\ref{Figure:Figure7}c. The average width of the emission components at 8.3 and 31.9\,GHz was 1.7 and 1.8\,ms, respectively. These observations are consistent with fan beam emission with an approximate width of $\pm$11$^{\circ}$ ($\pm$0.03 in phase units) at 8.3\,GHz and $\pm$7$^{\circ}$ ($\pm$0.02 in phase units) at 31.9\,GHz based on the widths of the pulse component distributions. Single pulse emission components were sometimes detected outside of these phase ranges during some rotations (e.g., at earlier phases compared to the pulse profile peak at 31.9\,GHz; see Figure~\ref{Figure:Figure7}c).

We found that XTE~J1810--197's pulse strength at 31.9\,GHz was significantly variable on timescales of $\sim$1000--4000\,s and often exhibited extended periods where pulsations were not detected (see Figure~\ref{Figure:Figure7}c; Methods). This behavior was intermittent and frequency- \\ dependent, as we did not see similar behavior during our simultaneous observations of the magnetar at 8.3\,GHz. This may be an intrinsic effect of the magnetar's emission mechanism at higher radio frequencies or caused by ISS.

The overall emission behavior at 8.3\,GHz is similar to the pulse morphology observed from the Galactic Center~(GC) magnetar, PSR~J1745--2900\citemain{Pearlman+2018}. However, in contrast to the GC~magnetar, XTE~J1810--197 shows a negligible amount of pulse broadening in its single pulse emission components at this frequency. During individual rotations, there is often variability in the frequency structure between the GC~magnetar's single pulse emission components\citemain{Pearlman+2018}, whereas the frequency structure in XTE~J1810--197's single pulses is typically uniform across all of the pulse components (e.g.,~see Figure~\ref{Figure:Figure8}a). In the case of XTE~J1810--197, the extent of these features ranged between $\sim$1--50\,MHz, which is smaller than the $\sim$100\,MHz frequency extent observed from the GC~magnetar\citemain{Pearlman+2018}. However, the 31.9\,GHz single pulses from XTE~J1810--197 did not show prominent evidence of this structure in any of its single pulse emission components. Similar spectral features have also been observed at lower frequencies (550--750\,MHz) in XTE~J1810--197's single pulse components\citemethods{Maan+2019}. This behavior is likely caused by ISS, but may also be intrinsic to the magnetar's emission mechanism.

If we assume that the frequency structure observed in the 8.3\,GHz single pulses is due to diffractive~ISS through a uniform Kolmogorov scattering medium, then the scintillation timescale ($\Delta t_{d}$) in seconds is given by\citemethods{Cordes+1998}:
\begin{equation}
\Delta t_{d}=A_{\text{ISS}}\frac{\sqrt{D\Delta\nu_{d}}}{V_{\text{ISS}}\nu},
\end{equation}
where $A_{\text{ISS}}$\,$=$\,2.53\,$\times$\,10$^{\text{4}}$\,km\,s$^{\text{--1}}$ is the ISS~velocity coefficient, $D$ is the distance to the source in kpc, $\Delta \nu_{d}$ is the scintillation bandwidth in MHz, $V_{\text{ISS}}$ is the ISS velocity, and $\nu$ is the observing frequency in GHz. Letting $V_{\text{ISS}}$ be equal to the pulsar's transverse velocity ($V_{\perp}$\,$=$\,212\,$\pm$\,35\,km\,s$^{\text{--1}}$)\citemethods{Helfand+2007}, $D$\,$=$\,3.5\,$\pm$\,0.5\,kpc\citemain{Minter+2008}, $\Delta \nu_{d}$\,$=$\,50\,MHz, and $\nu$\,$=$\,8.3\,GHz, we obtain a scintillation timescale of $\Delta t_{d}$\,$=$\,190\,$\pm$\,34\,s. The magnetar's radio pulse components displayed similar frequency structure over multiple consecutive rotations. The timescale over which we observed variations in the frequency structure is comparable to the predicted scintillation timescale. Additionally, the estimated pulse broadening timescale\citemethods{Cordes+1998} at 8.3\,GHz for a scintillation bandwidth of $\Delta \nu_{d}$\,$=$\,50\,MHz is:
\begin{equation}
\tau_{d}=\frac{C_{1}}{2\pi\Delta\nu_{d}}=\text{4\,ns}
\end{equation}
where $C_{1}$\,$=$\,1.16 for a uniform medium with a Kolmogorov wavenumber spectrum. The detection of such a pulse broadening magnitude is beyond the capability of our instrument.

\section{Comparisons with Fast Radio Bursts}
\label{Section:FastRadioBursts}

\indent Fast radio bursts~(FRBs) are bright, coherent pulses of radio emission with $\sim$$\mu$s--ms durations and fluences between roughly 0.01 and 1,000\,Jy\,ms (e.g., see refs.\,\citemethods{Petroff+2019, Cordes+2019} for recent reviews). They are thought to have extragalactic origins since their DMs exceed the values expected from Galactic free electrons along their lines of sight. Five FRBs have now been localized to host galaxies with redshifts between 0.034--0.66, which has established that their sources are located at extragalactic distances\citemethods{Chatterjee+2017, Bannister+2019, Ravi+2019, Prochaska+2019, Marcote+2020}. Thus far, over a hundred distinct FRB sources have been reported\citemethods{Petroff+2016}\textsuperscript{,}\customfootnote{6h}{~See http://frbcat.org.}. A wide variety of models, including cataclysmic and repeating scenarios, have been proposed to explain the progenitors of FRBs (e.g., see ref.\,\citemethods{Platts+2019}\textsuperscript{,}\customfootnote{7h}{~See http://frbtheorycat.org.} for a catalog). In particular, extragalactic magnetars have been suggested as one of the possible progenitor types (e.g., see refs.\,\citemain{Pearlman+2018, Bochenek+2020b}\textsuperscript{,}\citemethods{Pen+2015, Metzger+2017, Michilli+2018, Margalit+2018}).
	
Recently, a $\gtrsim$\,1.5\,MJy\,ms fluence radio burst was detected at 1.4\,GHz from the Galactic magnetar SGR~1935+2154 with the Survey for Transient Astronomical Radio Emission~2 (STARE2)\citemethods{Bochenek+2020a} during a period of enhanced X-ray activity\citemain{Bochenek+2020b}\textsuperscript{,}\citemethods{Palmer+2020, Younes+2020, Kennea+2020}. A less energetic radio burst, with a fluence of a few kJy\,ms between 400 and 800\,MHz, was also detected contemporaneously at lower frequencies using CHIME/FRB\citemethods{Scholz+2020}, along with a bright X-ray burst\citemain{Mereghetti+2020b}\textsuperscript{,}\citemethods{Mereghetti+2020a, Zhang+2020a}. We note that the lower apparent fluence of the radio burst detected by CHIME/FRB may be partially explained by the fact that the burst was detected in the telescope's sidelobe. The radio burst shown in ref.\,\citemethods{Scholz+2020} has two prominent emission components, which are separated by $\sim$30\,ms and have widths of $\sim$5\,ms. The dynamic spectrum of these emission components displays evidence of band-limited frequency structure that is variable between the two components\citemethods{Scholz+2020}.

Hereafter, we assume a distance of 12.5\,kpc to SGR~1935+2154 based on the magnetar's possible association with SNR~G57.2+0.8\citemethods{Kothes+2018}. However, we note that a range of distances between 4.5 and 12.5\,kpc have been suggested (e.g., see refs.\,\citemethods{Kothes+2018, Ranasinghe+2018, Zhou+2020}). The isotropic radio burst luminosity for a burst duration of $w$\,$=$\,1\,ms, based on the radio fluence reported in ref.\,\citemain{Bochenek+2020b}, is $L_{\nu}$\,$\approx$\,3\,$\times$\,10$^{\text{9}}$ Jy\,kpc$^{\text{2}}$\,$\approx$\,3\,$\times$\,10$^{\text{29}}$\,erg\,s$^{\text{--1}}$\,Hz$^{\text{--1}}$. This value far exceeds the luminosities of typical pulses from Galactic radio pulsars, rotating radio transients (RRATs), and giant radio pulses from Galactic pulsars, such as the Crab pulsar, by several orders of magnitude\citemethods{Keane2018}. The inferred brightness temperature, $T_{B}$, of SGR~1935+2154 at 1.4\,GHz during the time of the radio burst detected by STARE2 was $T_{B}$\,$\gtrsim$\,$\left(\frac{4}{\pi}\right)\frac{L_{\nu}}{2k_{\text{B}}(\nu w)^{\text{2}}}$ $\approx$\,7\,$\times$\,10$^{\text{32}}$\,K, where \\ $k_{\text{B}}$\,$\approx$\,1.38\,$\times$\,10$^{\text{--23}}$\,J\,K$^{\text{--1}}$ is Boltzmann's constant. This implies that the radio burst discovered by STARE2 would have been detected with a fluence of $\gtrsim$\,10\,mJy\,ms at a luminosity distance of 149\,Mpc (the luminosity distance of FRB~180916.J0158+65, which is currently the nearest localized FRB). Multiple radio bursts from FRB~180916.J0158+65 have already been detected with fluences above this level (e.g., see refs.\,\citemethods{CHIME+2019, CHIME+2020, Chawla+2020}). This suggests that active magnetars can generate radio bursts with enough energy to be detected from low redshift host galaxies. If the apparent brightness of such bursts are magnified via extrinsic propagation effects, such as plasma lensing\citemain{Pearlman+2018}\textsuperscript{,}\citemethods{Cordes+2017}, then extragalactic magnetars at high redshifts may also be a source of FRBs.

The radio emission from repeating FRB sources, such as FRB~121102 and FRB 180916.J0158+65, and the radio pulses detected from XTE~J1810--197 during its recent outburst have similar characteristics. Multicomponent radio pulses, with roughly millisecond widths, have been observed from both types of objects (e.g., see refs.\,\citemethods{Gajjar+2018, Hessels+2019, Majid+2020}). Significant frequency structure has also been observed in the radio pulses of both XTE~J1810--197\citemethods{Maan+2019} and the Galactic Center magnetar, PSR~J1745--2900\citemain{Pearlman+2018}, with a frequency extent that is similar to the spectral scales seen in bursts from repeating FRBs, such as FRB~121102\citemethods{Spitler+2016, Gajjar+2018, Michilli+2018}. We also found that some of XTE~J1810--197's radio pulse components were not detectable over a broad radio frequency range, which also resembles the behavior seen in radio bursts from repeating FRBs\citemethods{Law+2017, Majid+2020}.

While these similarities may indicate a common underlying emission mechanism between radio magnetars and FRBs, we note that there are several important differences between the radio emission from XTE~J1810--197 and FRBs. We did not find evidence that XTE~J1810--197's radio subpulses drifted downwards in frequency as time progressed. This ``sad trombone'' behavior is a characteristic feature in many radio bursts detected from repeating FRB sources (e.g., see refs.\,\citemethods{CHIME+2019, Hessels+2019}), but has thus far not been observed from radio magnetars. Additionally, the fluences of the 8.3 and 31.9\,GHz radio pulses from XTE~J1810--197 did not exceed $\sim$30\,Jy\,ms (see Figures~\ref{Figure:Figure7}b and~\ref{Figure:Figure7}c), which is incompatible with the energy output from FRB~180916.J0158+65. At a luminosity distance of 149\,Mpc, FRB~180916.J0158+65 produces radio bursts that are a factor of $\sim$10$^{\text{8}}$ more energetic than the radio single pulse emission from XTE~J1810--197. However, we note that a faint (30\,mJy) radio pulse was detected from SGR~1935+2154 approximately 2\,days after the arrival time of the $\gtrsim$\,1.5\,MJy\,ms radio burst\citemethods{Zhang+2020b}. The high fluence radio burst reported in ref.\,\citemain{Bochenek+2020b} may have been produced by energy release as a result of a fast reconnection event in the magnetar's over-twisted magnetosphere\citemethods{Lyutikov2002, Lyutikov+2020}. Less energetic radio bursts from magnetars, such as those from XTE~J1810--197 and SGR~1935+2154, may instead be powered through a combination of magnetic and rotational energy.

\newgeometry{top=1cm}

\begin{figure}
	\centering
	\includegraphics[trim=0cm 0cm 0cm 0cm, clip=false, scale=0.68, angle=0]{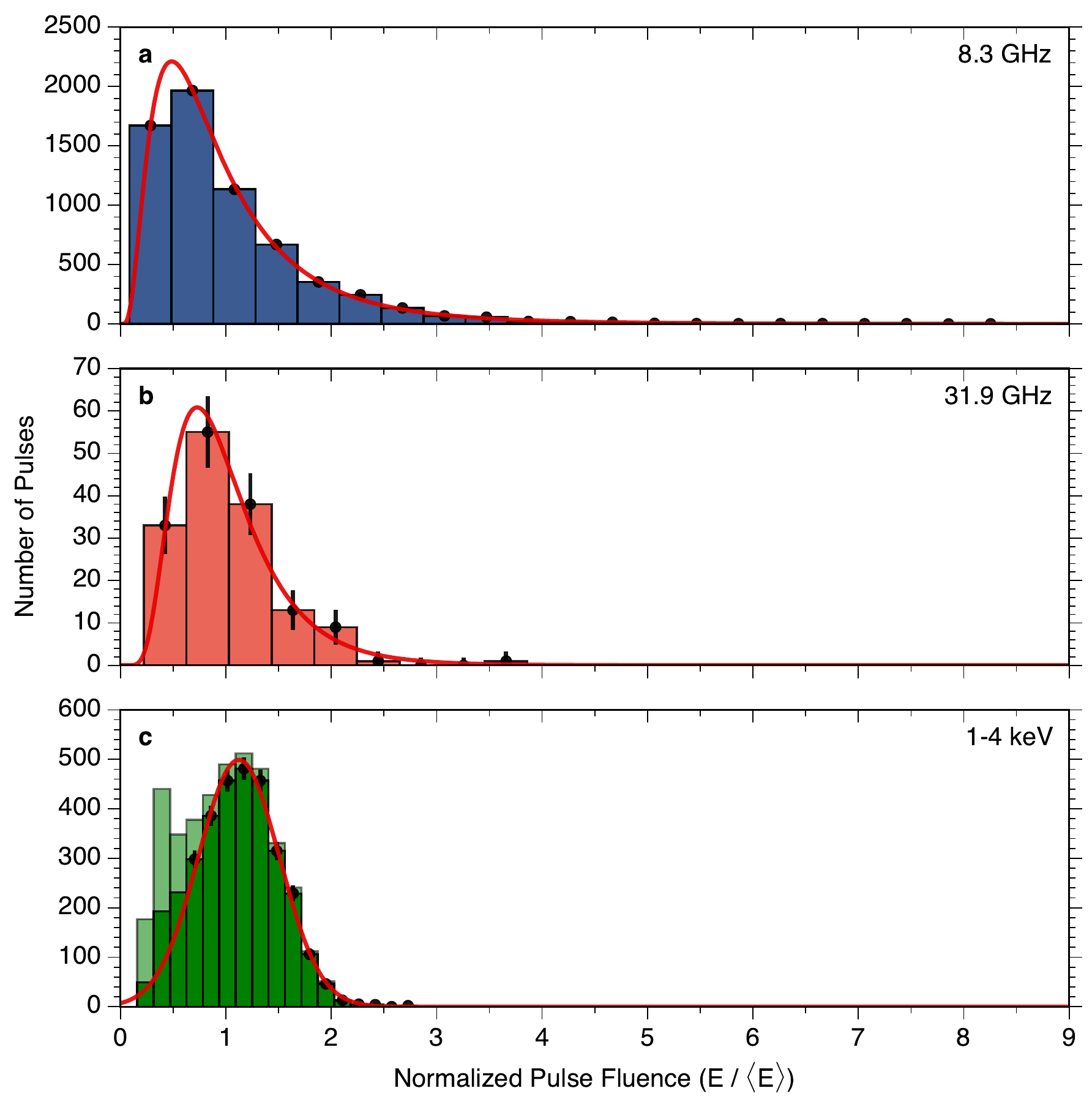}
	\caption{Pulse-energy distributions of the radio and X-ray pulses emitted by XTE~J1810--197. In panels~(a) and~(b), we show the distribution of (a)~8.3 and (b)~31.9\,GHz radio pulses detected with the DSN on MJD\,58530 above a S/N threshold of 7.0. Best-fit log-normal distributions are overlaid in red in both of these subpanels. The energy distribution of bright 1--4\,keV X-ray pulses detected by \textit{NICER} between MJDs\,58520 and 58540 is shown in panel~(c). The light green histogram corresponds to the total X-ray pulse distribution, and the dark green histogram shows the distribution derived from including only the brightest X-ray pulse during each rotation. The best-fit Gaussian distribution to the distribution of brightest X-ray pulses with fluences of $E/\langle E \rangle$\,$\gtrsim$\,0.6 is given by the red curve in panel~(c). The error bars shown in all of these panels are derived from Poisson statistics.}
	\label{Figure:Figure5}
\end{figure}

\begin{figure}
	\centering
	\includegraphics[trim=0cm 0cm 0cm 0cm, clip=false, scale=0.7, angle=0]{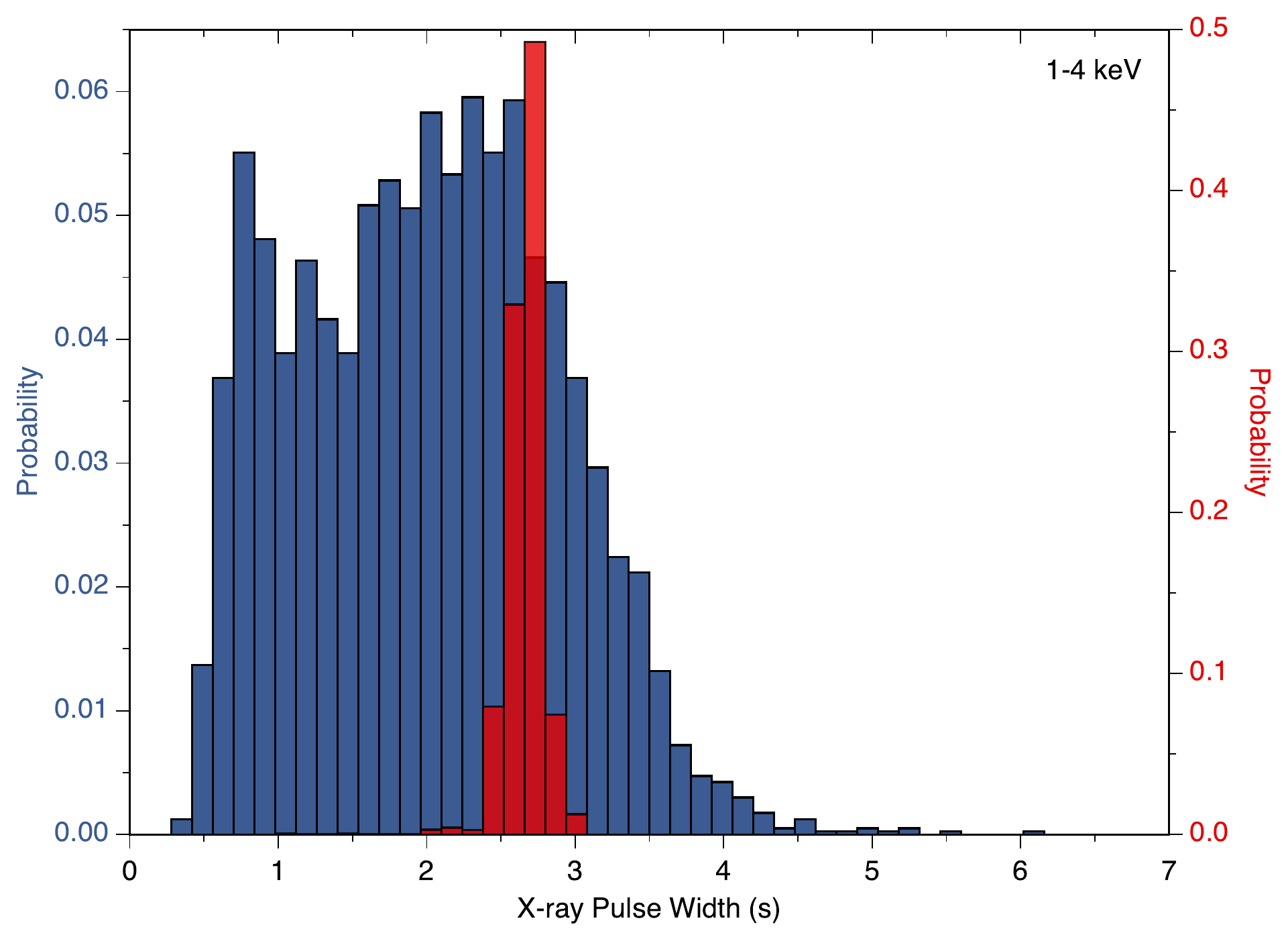}
	\caption{Distribution of X-ray pulse widths determined by the zero-crossing algorithm from \textit{NICER} observations between MJDs\,58520 and 58540 in the 1--4\,keV energy band. The observed pulse width distribution is shown in blue, and the simulated distribution derived from Monte Carlo realizations of the X-ray pulses is overlaid in red.}
	\label{Figure:Figure6}
\end{figure}

\begin{figure}
	\centering
	~~~~~~~~~~~~~\includegraphics[trim=0cm 0cm 0cm 0cm, clip=false, scale=0.7, angle=0]{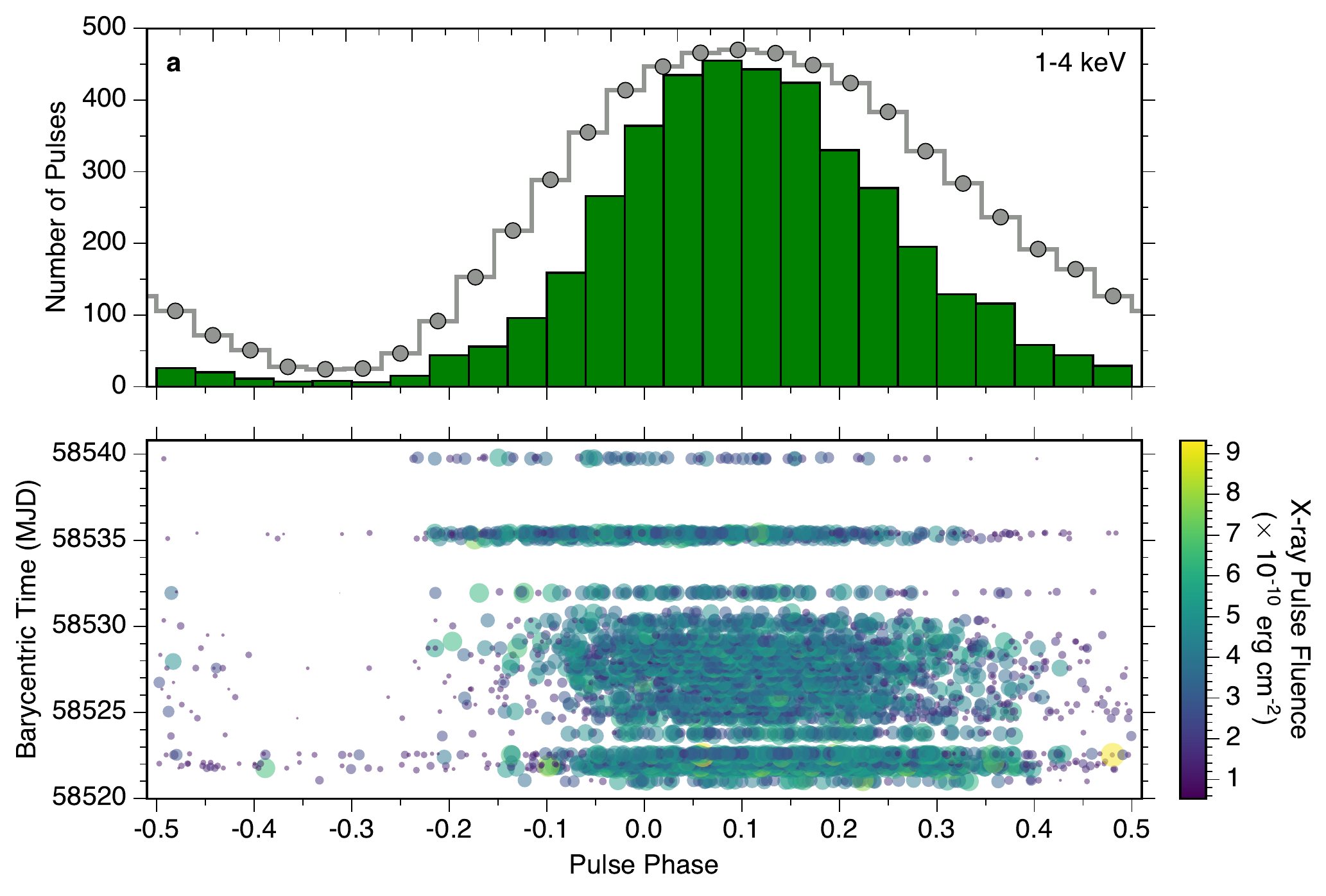}
	\\
	\includegraphics[trim=0cm 0cm 0cm 0cm, clip=false, scale=0.7, angle=0]{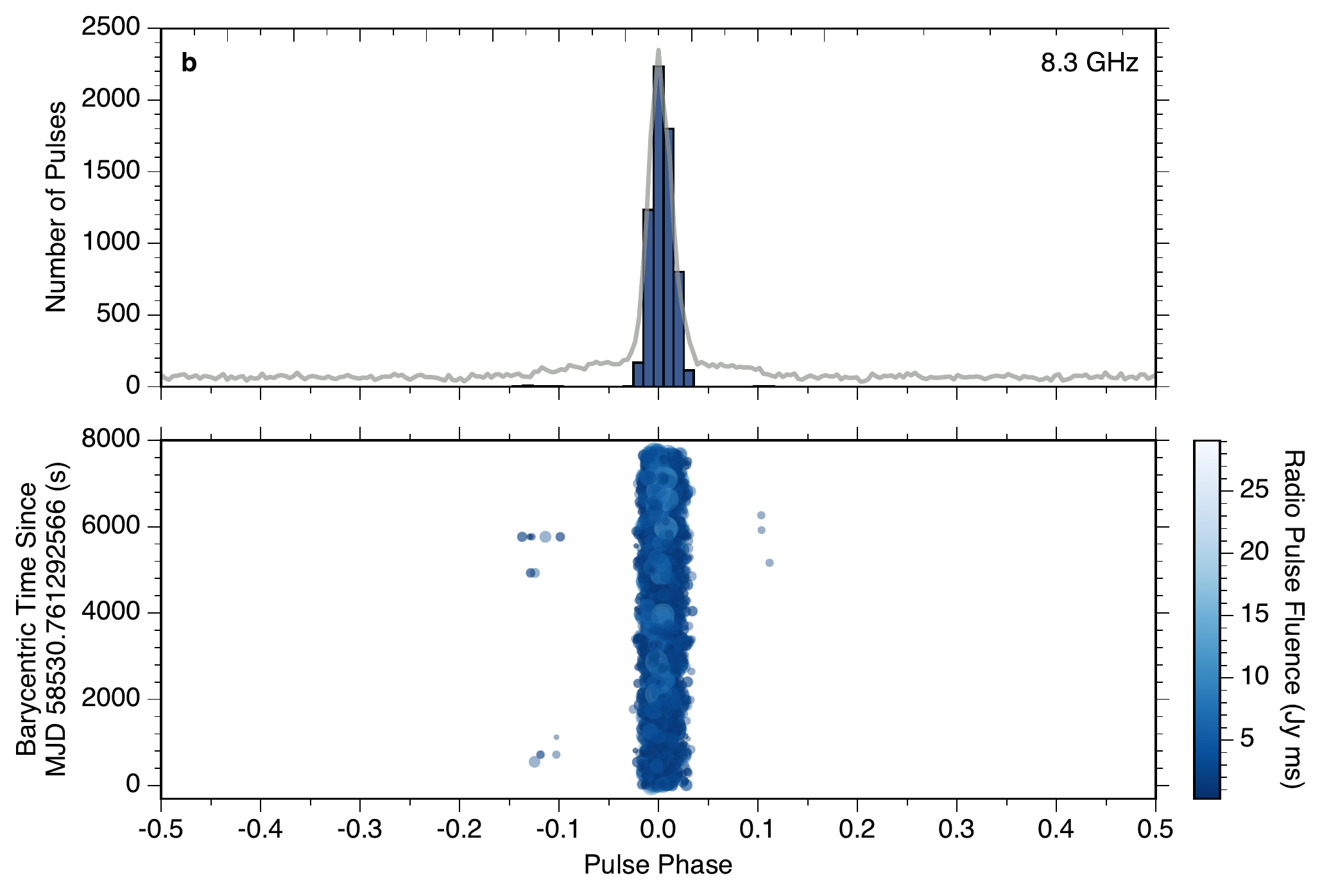}
\end{figure}

\begin{figure}
	\centering
	\includegraphics[trim=0cm 0cm 0cm 0cm, clip=false, scale=0.7, angle=0]{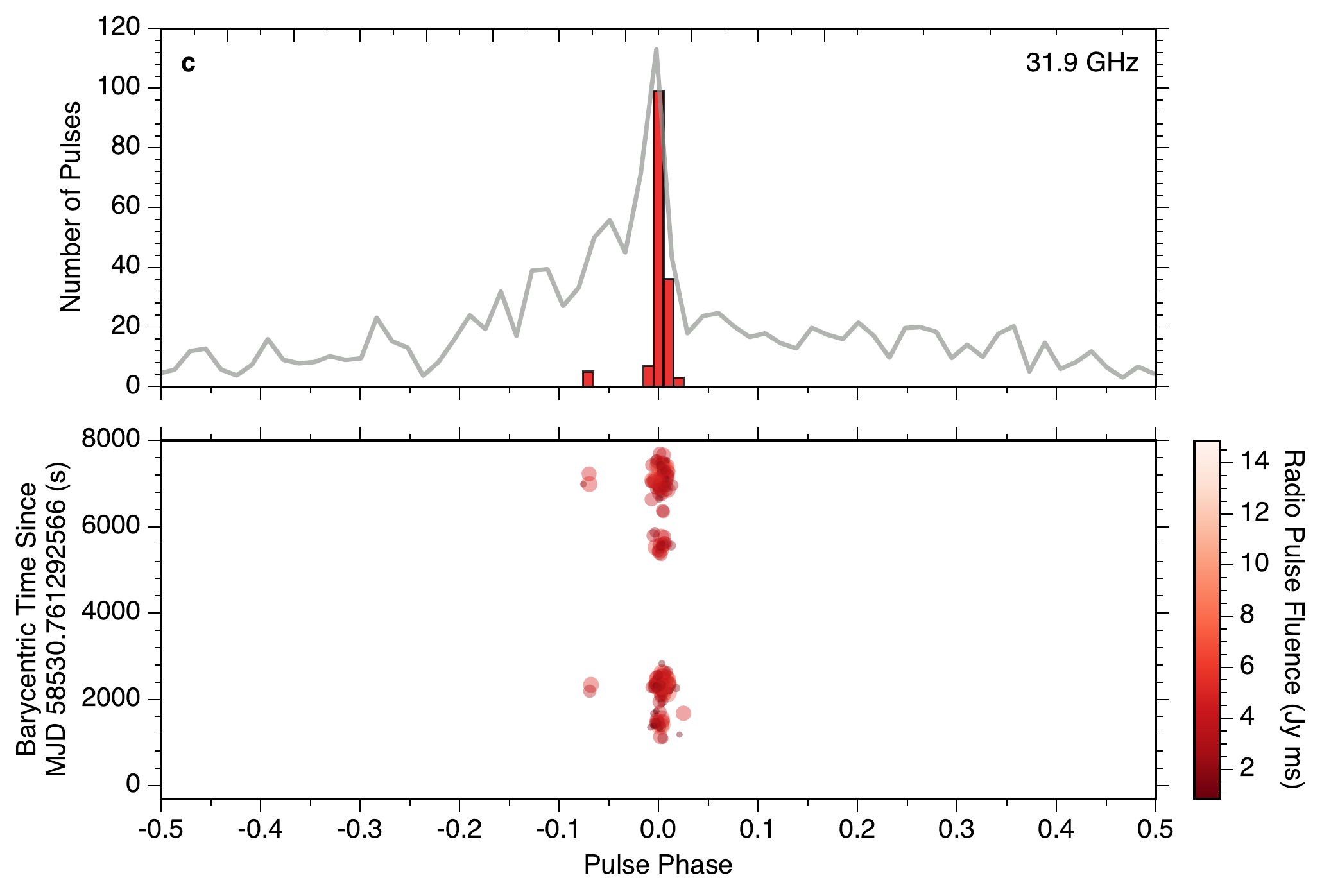}
	\caption{Distribution of X-ray and radio pulses as a function of time and XTE~J1810--197's rotational phase. In panel (a), we show the distribution of X-ray pulses detected in the 1--4\,keV energy band between MJDs\,58520 and 58540 with \textit{NICER}. The distribution of 8.3 and 31.9\,GHz radio pulses detected with S/N\,$\ge$\,7.0 on MJD\,58530 using the~DSN are shown in panels (b) and (c), respectively. The color and size of each data point in the bottom panels both indicate the fluence of each pulse. Histograms of the number of pulses as a function of pulse phase are provided in the top panels, and the bottom panels show their time-phase distribution. The folded pulse profiles are overlaid in gray in the top panels. The \textit{NICER} data shown in panel (a) are not continuous, unlike the radio observations, and gaps along the time axis in the bottom panel indicate times when \textit{NICER} was not observing the source.}
	\label{Figure:Figure7}
\end{figure}

\begin{figure}
	\centering
	\includegraphics[trim=0cm 0cm 0cm 0cm, clip=false, scale=0.55, angle=0]{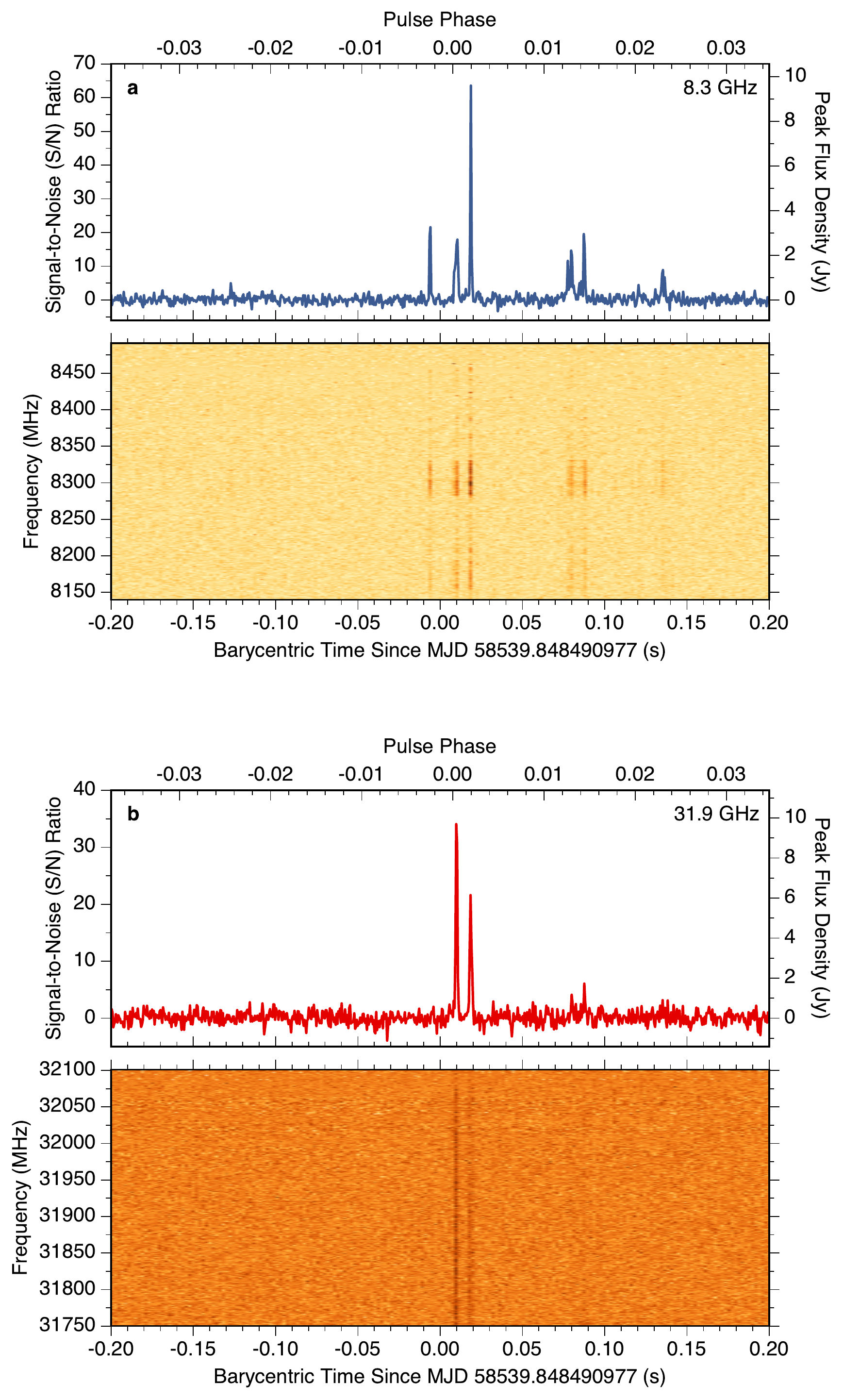}
	\caption{Example of bright radio pulses detected simultaneously at (a)~8.3 and (b)~31.9\,GHz from XTE~J1810--197 on MJD\,58539 during the same rotation of the neutron star. The top panels show the Stokes~I integrated single pulse profiles, and the Stokes~I dedispersed dynamic spectra are displayed in the bottom panels.}
	\label{Figure:Figure8}
\end{figure}

\newgeometry{top=2.5cm}

\section{Data Availability}

A catalog of the \textit{NICER} and DSN observations is provided in Table~\ref{Table:Table1}. The X-ray observations are accessible through the High Energy Astrophysics Science Archive Research Center (HEASARC) data archive\customfootnote{8h}{~See https://heasarc.gsfc.nasa.gov/cgi-bin/W3Browse/w3browse.pl.}. The radio data used in this paper are available from the corresponding author upon reasonable request.

\section{Code Availibility}

The corresponding author will provide the codes used to analyze the observations described in this paper upon reasonable request.

\newpage


\begin{table}
	\centering
	\caption{X-ray and Radio Observations of XTE~J1810--197}
	\medskip
	\begin{adjustbox}{width=1.0\textwidth}
		\begin{tabular}{cccccc}
			Instrument & Observation ID & Observation Start Time (UTC) & Observation Start Time (MJD) & Exposure Time (ks) & Count Rate$^{\text{a}}$ (counts\,s$^{\text{--1}}$) \\
			\hline
			\textit{NICER} & 1020420130 & 2019 Feb 06 23:59:20 & 58520.99954 & 4.39 & 48.9 \\
			\textit{NICER} & 1020420131 & 2019 Feb 08 00:29:56 & 58522.02079 & 4.83 & 48.5 \\
			\textit{NICER} & 1020420132 & 2019 Feb 09 15:06:20 & 58523.62940 & 1.09 & 47.1 \\
			\textit{NICER} & 1020420133 & 2019 Feb 10 05:01:40 & 58524.20949 & 0.79 & 45.9 \\
			\textit{NICER} & 1020420134 & 2019 Feb 11 01:07:00 & 58525.04653 & 2.00 & 46.6 \\
			\textit{NICER} & 1020420135 & 2019 Feb 12 07:04:20 & 58526.29468 & 1.28 & 46.0 \\
			\textit{NICER} & 1020420136 & 2019 Feb 13 02:42:41 & 58527.11297 & 2.75 & 46.1 \\
			\textit{NICER} & 1020420137 & 2019 Feb 14 08:08:00 & 58528.33889 & 2.13 & 46.1 \\
			\textit{NICER} & 1020420138 & 2019 Feb 15 02:46:28 & 58529.11560 & 1.50 & 45.5 \\
			\textit{NICER}$^{\dagger}$ & 1020420139 & 2019 Feb 16 00:28:00 & 58530.01944 & 1.04 & 45.5 \\
			DSN (DSS-34)$^{\dagger}$ &  & 2019 Feb 16 18:19:51 & 58530.76378 & 7.72 &  \\
			\textit{NICER} & 1020420140 & 2019 Feb 17 22:22:11 & 58531.93207 & 0.48 & 46.6 \\
			\textit{NICER} & 1020420141 & 2019 Feb 17 23:54:35 & 58531.99624 & 0.52 & 48.4 \\
			\textit{NICER} & 1020420142 & 2019 Feb 21 02:12:00 & 58535.09167 & 2.53 & 43.7 \\
			\textit{NICER}$^{\dagger}$ & 1020420143 & 2019 Feb 25 17:27:40 & 58539.72755 & 0.32 & 44.8 \\
			DSN (DSS-35)$^{\dagger}$ &  & 2019 Feb 25 18:37:09 & 58539.77580 & 6.71 &  \\
			
		\end{tabular}
	\end{adjustbox}
	
	\begin{minipage}{1.0\textwidth}
		~ \\
		\scriptsize $^{\text{a}}$ Average, non-background-subtracted X-ray count rate between 1 and 4\,keV. \\
		$^{\dagger}$ Data with simultaneous radio and X-ray observations. The overlapping radio and X-ray observations on 2019~February~16 and 2019~February~25 covered a total of $\sim$109\,s and $\sim$90\,s, respectively. \\
	\end{minipage}
	
	\label{Table:Table1}
\end{table}


\begin{table}
	\centering
	\caption{System Parameters of XTE~J1810--197}
	\medskip
	\begin{adjustbox}{width=0.7\textwidth}
		\setlength\extrarowheight{-5pt}
		\begin{tabular}{cc}
			Parameter & Value \\
			\hline
			\multicolumn{2}{c}{Fixed Values} \\
			\hline
			Right ascension (RA, J2000)\,$^{\text{ref.}}$\,\citemethods{Helfand+2007} & 18\,h\,09\,min\,51.08696\,s \\
			Declination (Dec, J2000)\,$^{\text{ref.}}$\,\citemethods{Helfand+2007} & --19$^{\circ}$\,43'\,51.9315'' \\
			Dispersion measure (DM)\,$^{\text{ref.}}$\,\citemain{Camilo+2006} & 178.0\,pc\,cm$^{\text{--3}}$ \\
			Distance ($D$)\,$^{\text{ref.}}$\,\citemain{Minter+2008} & 3.5\,kpc \\
			\hline
			\multicolumn{2}{c}{Measured Values} \\
			\hline
			Mean flux density ($S_{\nu}$) at 8.3 GHz$^{\text{a}}$ & 4.6\,$\pm$\,0.9\,mJy \\
			Mean flux density ($S_{\nu}$) at 31.9 GHz$^{\text{a}}$ & 3.7\,$\pm$\,0.7\,mJy \\
			Spectral index ($\alpha$) between 8.3 and 31.9 GHz$^{\text{a}}$ & --0.2\,$\pm$\,0.2 \\
			Mean flux density ($S_{\nu}$) at 8.3 GHz$^{\text{b}}$ & 4.2\,$\pm$\,0.8\,mJy \\
			Mean flux density ($S_{\nu}$) at 31.9 GHz$^{\text{b}}$ & 2.9\,$\pm$\,0.6\,mJy \\
			Spectral index ($\alpha$) between 8.3 and 31.9 GHz$^{\text{b}}$ & --0.3\,$\pm$\,0.2 \\
			\hline
			\multicolumn{2}{c}{Phase-coherent 8.3\,GHz timing solution on MJD\,58530.8} \\
			\hline
			Pulse frequency ($\nu$)$^{\text{c}}$ & 0.1804568(1)\,Hz \\
			Reference epoch (TDB)$^{\text{d}}$ & MJD\,58530.761334907 \\
			Observation span (TDB) & MJD\,58530.76--58530.85 \\
			Number of ToAs & 4 \\
			Solar system ephemeris & DE405 \\
			Timescale & TDB \\
			Weighted root-mean-square (RMS) residual & 2.4\,ms \\
			\hline
			\multicolumn{2}{c}{Phase-coherent 8.3\,GHz timing solution on MJD\,58539.8} \\
			\hline
			Pulse frequency ($\nu$)$^{\text{c}}$ & 0.18045638(5)\,Hz \\
			Reference epoch (TDB)$^{\text{d}}$ & MJD\,58539.774091226 \\
			Observation span (TDB) & MJD\,58539.77--58539.85 \\
			Number of ToAs & 4 \\
			Solar system ephemeris & DE405 \\
			Timescale & TDB \\
			Weighted root-mean-square (RMS) residual & 0.7\,ms
			
		\end{tabular}
		
	\end{adjustbox}
	
	\begin{minipage}{0.7\textwidth}
		~ \\
		\scriptsize $^{\text{a}}$ Measured value at T$_{\text{ref}}$\,$=$\,MJD\,58530.8. \\
		\scriptsize $^{\text{b}}$ Measured value at T$_{\text{ref}}$\,$=$\,MJD\,58539.8. \\
		\scriptsize $^{\text{c}}$ These values were derived by fitting for a constant rotational frequency, $\nu$. \\
		\scriptsize $^{\text{d}}$ Reference time corresponding to phase 0. \\
	\end{minipage}
	
	\label{Table:Table2}
\end{table}


\begin{table}
	\centering
	\caption{Relative Phase Shifts Between the X-ray Pulse Profiles of XTE~J1810--197}
	\medskip
	\begin{adjustbox}{width=0.5\textwidth}
		\begin{tabular}{cc}
			
			Energy Bands (keV) & Relative Phase Shift \\
			\hline
			(1--2)/(2--3) & 0.010\,$\pm$\,0.001 \\
			(2--3)/(3--4) & 0.012\,$\pm$\,0.001 \\
			(3--4)/(4--5) & --0.008\,$\pm$\,0.002 \\
			(4--5)/(5--10) & --0.025\,$\pm$\,0.003 \\
			(3--5)/(5--10) & --0.031\,$\pm$\,0.003 \\
			
		\end{tabular}
	\end{adjustbox}
	\\
	\label{Table:Table3}
\end{table}


\begin{table}
	\centering
	\caption{X-ray Pulsed Fractions of XTE~J1810--197}
	\medskip
	\begin{adjustbox}{width=0.5\textwidth}
		\begin{tabular}{cc}
			
			Energy Band (keV) & RMS Pulsed Fraction$^{\text{a}}$ \\
			\hline
			0.5--5 & 0.200\,$\pm$\,0.001 \\
			1--2 & 0.184\,$\pm$\,0.001 \\
			2--3 & 0.215\,$\pm$\,0.002 \\
			3--4 & 0.243\,$\pm$\,0.003 \\
			4--5 & 0.275\,$\pm$\,0.005 \\
			5--10 & 0.298\,$\pm$\,0.009 \\
			
		\end{tabular}
	\end{adjustbox}
	
	\begin{minipage}{0.5\textwidth}
		~ \\
		\scriptsize $^{\text{a}}$ Background-subtracted RMS pulsed fractions. \\
	\end{minipage}
	
	\label{Table:Table4}
\end{table}

\clearpage
\newpage

\end{methods}

\newpage

\bibliographystylemethods{naturemag}
\bibliographymethods{references}


\begin{thebibliography}{10}
\expandafter\ifx\csname url\endcsname\relax
  \def\url#1{\texttt{#1}}\fi
\expandafter\ifx\csname urlprefix\endcsname\relax\def\urlprefix{URL }\fi
\providecommand{\bibinfo}[2]{#2}
\providecommand{\eprint}[2][]{\url{#2}}

\bibitem{Olausen+2014}
\bibinfo{author}{{Olausen}, S.~A.} \& \bibinfo{author}{{Kaspi}, V.~M.}
\newblock \bibinfo{title}{{The McGill Magnetar Catalog}}.
\newblock \emph{\bibinfo{journal}{ApJS}} \textbf{\bibinfo{volume}{212}},
  \bibinfo{pages}{6} (\bibinfo{year}{2014}).
\newblock \eprint{1309.4167}.

\bibitem{Kaspi+2017}
\bibinfo{author}{{Kaspi}, V.~M.} \& \bibinfo{author}{{Beloborodov}, A.~M.}
\newblock \bibinfo{title}{{Magnetars}}.
\newblock \emph{\bibinfo{journal}{Ann. Rev. Astron. Astrophys.}}
  \textbf{\bibinfo{volume}{55}}, \bibinfo{pages}{261--301}
  (\bibinfo{year}{2017}).
\newblock \eprint{1703.00068}.

\bibitem{Camilo+2006}
\bibinfo{author}{{Camilo}, F.} \emph{et~al.}
\newblock \bibinfo{title}{{Transient pulsed radio emission from a magnetar}}.
\newblock \emph{\bibinfo{journal}{Nature}} \textbf{\bibinfo{volume}{442}},
  \bibinfo{pages}{892--895} (\bibinfo{year}{2006}).
\newblock \eprint{astro-ph/0605429}.

\bibitem{Camilo+2007b}
\bibinfo{author}{{Camilo}, F.}, \bibinfo{author}{{Ransom}, S.~M.},
  \bibinfo{author}{{Halpern}, J.~P.} \& \bibinfo{author}{{Reynolds}, J.}
\newblock \bibinfo{title}{{1E 1547.0-5408: A Radio-emitting Magnetar with a
  Rotation Period of 2 Seconds}}.
\newblock \emph{\bibinfo{journal}{ApJL}} \textbf{\bibinfo{volume}{666}},
  \bibinfo{pages}{L93--L96} (\bibinfo{year}{2007}).
\newblock \eprint{0708.0002}.

\bibitem{Pearlman+2018}
\bibinfo{author}{{Pearlman}, A.~B.}, \bibinfo{author}{{Majid}, W.~A.},
  \bibinfo{author}{{Prince}, T.~A.}, \bibinfo{author}{{Kocz}, J.} \&
  \bibinfo{author}{{Horiuchi}, S.}
\newblock \bibinfo{title}{{Pulse Morphology of the Galactic Center Magnetar PSR
  J1745-2900}}.
\newblock \emph{\bibinfo{journal}{ApJ}} \textbf{\bibinfo{volume}{866}},
  \bibinfo{pages}{160} (\bibinfo{year}{2018}).
\newblock \eprint{1809.02140}.

\bibitem{Levin+2010}
\bibinfo{author}{{Levin}, L.} \emph{et~al.}
\newblock \bibinfo{title}{{A Radio-loud Magnetar in X-ray Quiescence}}.
\newblock \emph{\bibinfo{journal}{ApJL}} \textbf{\bibinfo{volume}{721}},
  \bibinfo{pages}{L33--L37} (\bibinfo{year}{2010}).
\newblock \eprint{1007.1052}.

\bibitem{Lyne+2018}
\bibinfo{author}{{Lyne}, A.} \emph{et~al.}
\newblock \bibinfo{title}{{Intense radio flare from the magnetar XTE
  J1810-197}}.
\newblock \emph{\bibinfo{journal}{The Astronomer's Telegram}}
  \textbf{\bibinfo{volume}{12284}} (\bibinfo{year}{2018}).

\bibitem{Hurley+1999}
\bibinfo{author}{{Hurley}, K.} \emph{et~al.}
\newblock \bibinfo{title}{{A giant periodic flare from the soft {$\gamma$}-ray
  repeater SGR1900+14}}.
\newblock \emph{\bibinfo{journal}{Nature}} \textbf{\bibinfo{volume}{397}},
  \bibinfo{pages}{41--43} (\bibinfo{year}{1999}).
\newblock \eprint{astro-ph/9811443}.

\bibitem{Palmer+2005}
\bibinfo{author}{{Palmer}, D.~M.} \emph{et~al.}
\newblock \bibinfo{title}{{A giant {$\gamma$}-ray flare from the magnetar SGR
  1806 - 20}}.
\newblock \emph{\bibinfo{journal}{Nature}} \textbf{\bibinfo{volume}{434}},
  \bibinfo{pages}{1107--1109} (\bibinfo{year}{2005}).
\newblock \eprint{astro-ph/0503030}.

\bibitem{Woods+2005}
\bibinfo{author}{{Woods}, P.~M.} \emph{et~al.}
\newblock \bibinfo{title}{{X-Ray Bursts from the Transient Magnetar Candidate
  XTE J1810-197}}.
\newblock \emph{\bibinfo{journal}{ApJ}} \textbf{\bibinfo{volume}{629}},
  \bibinfo{pages}{985--997} (\bibinfo{year}{2005}).
\newblock \eprint{astro-ph/0505039}.

\bibitem{Bochenek+2020b}
\bibinfo{author}{{Bochenek}, C.} \emph{et~al.}
\newblock \bibinfo{title}{{Independent detection of the radio burst reported in
  ATel \#13681 with STARE2}}.
\newblock \emph{\bibinfo{journal}{The Astronomer's Telegram}}
  \textbf{\bibinfo{volume}{13684}} (\bibinfo{year}{2020}).

\bibitem{Ibrahim+2004}
\bibinfo{author}{{Ibrahim}, A.~I.} \emph{et~al.}
\newblock \bibinfo{title}{{Discovery of a Transient Magnetar: XTE J1810-197}}.
\newblock \emph{\bibinfo{journal}{ApJL}} \textbf{\bibinfo{volume}{609}},
  \bibinfo{pages}{L21--L24} (\bibinfo{year}{2004}).
\newblock \eprint{astro-ph/0310665}.

\bibitem{Camilo+2016}
\bibinfo{author}{{Camilo}, F.} \emph{et~al.}
\newblock \bibinfo{title}{{Radio Disappearance of the Magnetar XTE J1810-197
  and Continued X-ray Timing}}.
\newblock \emph{\bibinfo{journal}{ApJ}} \textbf{\bibinfo{volume}{820}},
  \bibinfo{pages}{110} (\bibinfo{year}{2016}).
\newblock \eprint{1603.02170}.

\bibitem{Minter+2008}
\bibinfo{author}{{Minter}, A.~H.}, \bibinfo{author}{{Camilo}, F.},
  \bibinfo{author}{{Ransom}, S.~M.}, \bibinfo{author}{{Halpern}, J.~P.} \&
  \bibinfo{author}{{Zimmerman}, N.}
\newblock \bibinfo{title}{{Neutral Hydrogen Absorption toward XTE J1810-197:
  The Distance to a Radio-emitting Magnetar}}.
\newblock \emph{\bibinfo{journal}{ApJ}} \textbf{\bibinfo{volume}{676}},
  \bibinfo{pages}{1189--1199} (\bibinfo{year}{2008}).
\newblock \eprint{0705.4403}.

\bibitem{Camilo+2007c}
\bibinfo{author}{{Camilo}, F.} \emph{et~al.}
\newblock \bibinfo{title}{{The Variable Radio-to-X-Ray Spectrum of the Magnetar
  XTE J1810-197}}.
\newblock \emph{\bibinfo{journal}{ApJ}} \textbf{\bibinfo{volume}{669}},
  \bibinfo{pages}{561--569} (\bibinfo{year}{2007}).
\newblock \eprint{0705.4095}.

\bibitem{Pintore+2019}
\bibinfo{author}{{Pintore}, F.} \emph{et~al.}
\newblock \bibinfo{title}{{The 11 yr of low activity of the magnetar XTE
  J1810-197}}.
\newblock \emph{\bibinfo{journal}{MNRAS}} \textbf{\bibinfo{volume}{483}},
  \bibinfo{pages}{3832--3838} (\bibinfo{year}{2019}).
\newblock \eprint{1812.03153}.

\bibitem{Majid+2019}
\bibinfo{author}{{Majid}, W.~A.} \emph{et~al.}
\newblock \bibinfo{title}{{High Frequency Radio Observations of the Reactivated
  Magnetar XTE J1810-197}}.
\newblock \emph{\bibinfo{journal}{The Astronomer's Telegram}}
  \textbf{\bibinfo{volume}{12353}} (\bibinfo{year}{2019}).

\bibitem{Dai+2019}
\bibinfo{author}{{Dai}, S.} \emph{et~al.}
\newblock \bibinfo{title}{{Wideband Polarized Radio Emission from the Newly
  Revived Magnetar XTE J1810-197}}.
\newblock \emph{\bibinfo{journal}{ApJL}} \textbf{\bibinfo{volume}{874}},
  \bibinfo{pages}{L14} (\bibinfo{year}{2019}).
\newblock \eprint{1902.04689}.

\bibitem{Gotthelf+2019}
\bibinfo{author}{{Gotthelf}, E.~V.} \emph{et~al.}
\newblock \bibinfo{title}{{The 2018 X-Ray and Radio Outburst of Magnetar XTE
  J1810-197}}.
\newblock \emph{\bibinfo{journal}{ApJL}} \textbf{\bibinfo{volume}{874}},
  \bibinfo{pages}{L25} (\bibinfo{year}{2019}).
\newblock \eprint{1902.08358}.

\bibitem{Guver+2019}
\bibinfo{author}{{G{\"u}ver}, T.} \emph{et~al.}
\newblock \bibinfo{title}{{NICER Observations of the 2018 Outburst of XTE
  J1810{\ensuremath{-}}197}}.
\newblock \emph{\bibinfo{journal}{ApJL}} \textbf{\bibinfo{volume}{877}},
  \bibinfo{pages}{L30} (\bibinfo{year}{2019}).
\newblock \eprint{1905.04440}.

\bibitem{Pearlman+2019}
\bibinfo{author}{{Pearlman}, A.~B.}, \bibinfo{author}{{Majid}, W.~A.} \&
  \bibinfo{author}{{Prince}, T.~A.}
\newblock \bibinfo{title}{{Observations of Radio Magnetars with the Deep Space
  Network}}.
\newblock \emph{\bibinfo{journal}{Advances in Astronomy}}
  \textbf{\bibinfo{volume}{2019}}, \bibinfo{pages}{6325183}
  (\bibinfo{year}{2019}).
\newblock \eprint{1902.10712}.

\bibitem{Gotthelf+2005}
\bibinfo{author}{{Gotthelf}, E.~V.} \& \bibinfo{author}{{Halpern}, J.~P.}
\newblock \bibinfo{title}{{The Spectral Evolution of Transient Anomalous X-Ray
  Pulsar XTE J1810-197}}.
\newblock \emph{\bibinfo{journal}{ApJ}} \textbf{\bibinfo{volume}{632}},
  \bibinfo{pages}{1075--1085} (\bibinfo{year}{2005}).
\newblock \eprint{astro-ph/0506511}.

\bibitem{Beloborodov+2016}
\bibinfo{author}{{Beloborodov}, A.~M.} \& \bibinfo{author}{{Li}, X.}
\newblock \bibinfo{title}{{Magnetar Heating}}.
\newblock \emph{\bibinfo{journal}{ApJ}} \textbf{\bibinfo{volume}{833}},
  \bibinfo{pages}{261} (\bibinfo{year}{2016}).
\newblock \eprint{1605.09077}.

\bibitem{Ozel+2013}
\bibinfo{author}{{{\"O}zel}, F.}
\newblock \bibinfo{title}{{Surface emission from neutron stars and implications
  for the physics of their interiors}}.
\newblock \emph{\bibinfo{journal}{Reports on Progress in Physics}}
  \textbf{\bibinfo{volume}{76}}, \bibinfo{pages}{016901}
  (\bibinfo{year}{2013}).
\newblock \eprint{1210.0916}.

\bibitem{Perna+2008}
\bibinfo{author}{{Perna}, R.} \& \bibinfo{author}{{Gotthelf}, E.~V.}
\newblock \bibinfo{title}{{Constraints on the Emission and Viewing Geometry of
  the Transient Anomalous X-Ray Pulsar XTE J1810-197}}.
\newblock \emph{\bibinfo{journal}{ApJ}} \textbf{\bibinfo{volume}{681}},
  \bibinfo{pages}{522--529} (\bibinfo{year}{2008}).
\newblock \eprint{0803.2042}.

\bibitem{Camilo+2007a}
\bibinfo{author}{{Camilo}, F.} \emph{et~al.}
\newblock \bibinfo{title}{{The Magnetar XTE J1810-197: Variations in Torque,
  Radio Flux Density, and Pulse Profile Morphology}}.
\newblock \emph{\bibinfo{journal}{ApJ}} \textbf{\bibinfo{volume}{663}},
  \bibinfo{pages}{497--504} (\bibinfo{year}{2007}).
\newblock \eprint{astro-ph/0610685}.

\bibitem{Beloborodov2009}
\bibinfo{author}{{Beloborodov}, A.~M.}
\newblock \bibinfo{title}{{Untwisting Magnetospheres of Neutron Stars}}.
\newblock \emph{\bibinfo{journal}{ApJ}} \textbf{\bibinfo{volume}{703}},
  \bibinfo{pages}{1044--1060} (\bibinfo{year}{2009}).
\newblock \eprint{0812.4873}.

\bibitem{Archibald+2017}
\bibinfo{author}{{Archibald}, R.~F.} \emph{et~al.}
\newblock \bibinfo{title}{{Magnetar-like X-Ray Bursts Suppress Pulsar Radio
  Emission}}.
\newblock \emph{\bibinfo{journal}{ApJL}} \textbf{\bibinfo{volume}{849}},
  \bibinfo{pages}{L20} (\bibinfo{year}{2017}).
\newblock \eprint{1710.03718}.

\bibitem{Yamasaki+2019}
\bibinfo{author}{{Yamasaki}, S.}, \bibinfo{author}{{Kisaka}, S.},
  \bibinfo{author}{{Terasawa}, T.} \& \bibinfo{author}{{Enoto}, T.}
\newblock \bibinfo{title}{{Relativistic fireball reprise: radio suppression at
  the onset of short magnetar bursts}}.
\newblock \emph{\bibinfo{journal}{MNRAS}} \textbf{\bibinfo{volume}{483}},
  \bibinfo{pages}{4175--4186} (\bibinfo{year}{2019}).
\newblock \eprint{1810.06353}.

\bibitem{Mereghetti+2020b}
\bibinfo{author}{{Mereghetti}, S.} \emph{et~al.}
\newblock \bibinfo{title}{{INTEGRAL discovery of a burst with associated radio
  emission from the magnetar SGR 1935+2154}}.
\newblock \emph{\bibinfo{journal}{arXiv e-prints}}
  \bibinfo{pages}{arXiv:2005.06335} (\bibinfo{year}{2020}).
\newblock \eprint{2005.06335}.

\end{thebibliography}


\begin{thebibliography}{10}
\expandafter\ifx\csname url\endcsname\relax
  \def\url#1{\texttt{#1}}\fi
\expandafter\ifx\csname urlprefix\endcsname\relax\def\urlprefix{URL }\fi
\providecommand{\bibinfo}[2]{#2}
\providecommand{\eprint}[2][]{\url{#2}}

\bibitem{Gendreau+2016}
\bibinfo{author}{{Gendreau}, K.~C.} \emph{et~al.}
\newblock \bibinfo{title}{{The Neutron star Interior Composition Explorer
  (NICER): design and development}}.
\newblock In \emph{\bibinfo{booktitle}{Space Telescopes and Instrumentation
  2016: Ultraviolet to Gamma Ray}}, vol. \bibinfo{volume}{9905} of
  \emph{\bibinfo{series}{Proc. Soc. Photo-Opt. Instrum. Eng.}},
  \bibinfo{pages}{99051H} (\bibinfo{year}{2016}).

\bibitem{Blackburn1995}
\bibinfo{author}{{Blackburn}, J.~K.}
\newblock \bibinfo{title}{{FTOOLS: A FITS Data Processing and Analysis Software
  Package}}.
\newblock In \bibinfo{editor}{{Shaw}, R.~A.}, \bibinfo{editor}{{Payne}, H.~E.}
  \& \bibinfo{editor}{{Hayes}, J.~J.~E.} (eds.)
  \emph{\bibinfo{booktitle}{Astronomical Data Analysis Software and Systems
  IV}}, vol.~\bibinfo{volume}{77} of \emph{\bibinfo{series}{Astronomical
  Society of the Pacific Conference Series}}, \bibinfo{pages}{367}
  (\bibinfo{year}{1995}).

\bibitem{Ransom2001}
\bibinfo{author}{{Ransom}, S.~M.}
\newblock \emph{\bibinfo{title}{{New search techniques for binary pulsars}}}.
\newblock Ph.D. thesis, \bibinfo{school}{Harvard University}
  (\bibinfo{year}{2001}).

\bibitem{Taylor1992}
\bibinfo{author}{{Taylor}, J.~H.}
\newblock \bibinfo{title}{{Pulsar Timing and Relativistic Gravity}}.
\newblock \emph{\bibinfo{journal}{Philosophical Transactions of the Royal
  Society of London Series A}} \textbf{\bibinfo{volume}{341}},
  \bibinfo{pages}{117--134} (\bibinfo{year}{1992}).

\bibitem{Hobbs+2006}
\bibinfo{author}{{Hobbs}, G.~B.}, \bibinfo{author}{{Edwards}, R.~T.} \&
  \bibinfo{author}{{Manchester}, R.~N.}
\newblock \bibinfo{title}{{TEMPO2, a new pulsar-timing package - I. An
  overview}}.
\newblock \emph{\bibinfo{journal}{MNRAS}} \textbf{\bibinfo{volume}{369}},
  \bibinfo{pages}{655--672} (\bibinfo{year}{2006}).
\newblock \eprint{astro-ph/0603381}.

\bibitem{Helfand+2007}
\bibinfo{author}{{Helfand}, D.~J.} \emph{et~al.}
\newblock \bibinfo{title}{{VLBA Measurement of the Transverse Velocity of the
  Magnetar XTE J1810-197}}.
\newblock \emph{\bibinfo{journal}{ApJ}} \textbf{\bibinfo{volume}{662}},
  \bibinfo{pages}{1198--1203} (\bibinfo{year}{2007}).
\newblock \eprint{astro-ph/0703336}.

\bibitem{Scholz+2017}
\bibinfo{author}{{Scholz}, P.} \emph{et~al.}
\newblock \bibinfo{title}{{Spin-down Evolution and Radio Disappearance of the
  Magnetar PSR J1622-4950}}.
\newblock \emph{\bibinfo{journal}{ApJ}} \textbf{\bibinfo{volume}{841}},
  \bibinfo{pages}{126} (\bibinfo{year}{2017}).
\newblock \eprint{1705.04899}.

\bibitem{Abdo+2010}
\bibinfo{author}{{Abdo}, A.~A.} \emph{et~al.}
\newblock \bibinfo{title}{{Fermi Large Area Telescope Observations of the Crab
  Pulsar And Nebula}}.
\newblock \emph{\bibinfo{journal}{ApJ}} \textbf{\bibinfo{volume}{708}},
  \bibinfo{pages}{1254--1267} (\bibinfo{year}{2010}).
\newblock \eprint{0911.2412}.

\bibitem{Levin+2019}
\bibinfo{author}{{Levin}, L.} \emph{et~al.}
\newblock \bibinfo{title}{{Spin frequency evolution and pulse profile
  variations of the recently re-activated radio magnetar XTE J1810-197}}.
\newblock \emph{\bibinfo{journal}{MNRAS}} \textbf{\bibinfo{volume}{488}},
  \bibinfo{pages}{5251--5258} (\bibinfo{year}{2019}).
\newblock \eprint{1903.02660}.

\bibitem{Dib+2012}
\bibinfo{author}{{Dib}, R.}, \bibinfo{author}{{Kaspi}, V.~M.},
  \bibinfo{author}{{Scholz}, P.} \& \bibinfo{author}{{Gavriil}, F.~P.}
\newblock \bibinfo{title}{{RXTE Observations of Anomalous X-Ray Pulsar 1E
  1547.0-5408 during and after its 2008 and 2009 Outbursts}}.
\newblock \emph{\bibinfo{journal}{ApJ}} \textbf{\bibinfo{volume}{748}},
  \bibinfo{pages}{3} (\bibinfo{year}{2012}).
\newblock \eprint{1201.2668}.

\bibitem{Camilo+2018}
\bibinfo{author}{{Camilo}, F.} \emph{et~al.}
\newblock \bibinfo{title}{{Revival of the Magnetar PSR J1622-4950: Observations
  with MeerKAT, Parkes, XMM-Newton, Swift, Chandra, and NuSTAR}}.
\newblock \emph{\bibinfo{journal}{ApJ}} \textbf{\bibinfo{volume}{856}},
  \bibinfo{pages}{180} (\bibinfo{year}{2018}).
\newblock \eprint{1804.01933}.

\bibitem{Enoto+2019}
\bibinfo{author}{Enoto, T.}, \bibinfo{author}{Kisaka, S.} \&
  \bibinfo{author}{Shibata, S.}
\newblock \bibinfo{title}{Observational diversity of magnetized neutron stars}.
\newblock \emph{\bibinfo{journal}{Reports on Progress in Physics}}
  \textbf{\bibinfo{volume}{82}}, \bibinfo{pages}{106901}
  (\bibinfo{year}{2019}).
\newblock \urlprefix\url{https://doi.org/10.1088%2F1361-6633%2Fab3def}.

\bibitem{McLaughlin+2003}
\bibinfo{author}{{McLaughlin}, M.~A.} \& \bibinfo{author}{{Cordes}, J.~M.}
\newblock \bibinfo{title}{{Searches for Giant Pulses from Extragalactic
  Pulsars}}.
\newblock \emph{\bibinfo{journal}{ApJ}} \textbf{\bibinfo{volume}{596}},
  \bibinfo{pages}{982--996} (\bibinfo{year}{2003}).
\newblock \eprint{astro-ph/0304365}.

\bibitem{Serylak+2009}
\bibinfo{author}{{Serylak}, M.} \emph{et~al.}
\newblock \bibinfo{title}{{Simultaneous multifrequency single-pulse properties
  of AXP XTE J1810-197}}.
\newblock \emph{\bibinfo{journal}{MNRAS}} \textbf{\bibinfo{volume}{394}},
  \bibinfo{pages}{295--308} (\bibinfo{year}{2009}).
\newblock \eprint{0811.3829}.

\bibitem{Maan+2019}
\bibinfo{author}{{Maan}, Y.}, \bibinfo{author}{{Joshi}, B.~C.},
  \bibinfo{author}{{Surnis}, M.~P.}, \bibinfo{author}{{Bagchi}, M.} \&
  \bibinfo{author}{{Manoharan}, P.~K.}
\newblock \bibinfo{title}{{Distinct properties of the radio burst emission from
  the magnetar XTE J1810-197}}.
\newblock \emph{\bibinfo{journal}{ApJL}} \textbf{\bibinfo{volume}{882}},
  \bibinfo{pages}{L9} (\bibinfo{year}{2019}).
\newblock \eprint{1908.04304}.

\bibitem{Cordes+1998}
\bibinfo{author}{{Cordes}, J.~M.} \& \bibinfo{author}{{Rickett}, B.~J.}
\newblock \bibinfo{title}{{Diffractive Interstellar Scintillation Timescales
  and Velocities}}.
\newblock \emph{\bibinfo{journal}{ApJ}} \textbf{\bibinfo{volume}{507}},
  \bibinfo{pages}{846--860} (\bibinfo{year}{1998}).

\bibitem{Petroff+2019}
\bibinfo{author}{{Petroff}, E.}, \bibinfo{author}{{Hessels}, J.~W.~T.} \&
  \bibinfo{author}{{Lorimer}, D.~R.}
\newblock \bibinfo{title}{{Fast radio bursts}}.
\newblock \emph{\bibinfo{journal}{A\&A Review}} \textbf{\bibinfo{volume}{27}},
  \bibinfo{pages}{4} (\bibinfo{year}{2019}).
\newblock \eprint{1904.07947}.

\bibitem{Cordes+2019}
\bibinfo{author}{{Cordes}, J.~M.} \& \bibinfo{author}{{Chatterjee}, S.}
\newblock \bibinfo{title}{{Fast Radio Bursts: An Extragalactic Enigma}}.
\newblock \emph{\bibinfo{journal}{Ann. Rev. Astron. Astrophys.}}
  \textbf{\bibinfo{volume}{57}}, \bibinfo{pages}{417--465}
  (\bibinfo{year}{2019}).
\newblock \eprint{1906.05878}.

\bibitem{Chatterjee+2017}
\bibinfo{author}{{Chatterjee}, S.} \emph{et~al.}
\newblock \bibinfo{title}{{A direct localization of a fast radio burst and its
  host}}.
\newblock \emph{\bibinfo{journal}{Nature}} \textbf{\bibinfo{volume}{541}},
  \bibinfo{pages}{58--61} (\bibinfo{year}{2017}).
\newblock \eprint{1701.01098}.

\bibitem{Bannister+2019}
\bibinfo{author}{{Bannister}, K.~W.} \emph{et~al.}
\newblock \bibinfo{title}{{A single fast radio burst localized to a massive
  galaxy at cosmological distance}}.
\newblock \emph{\bibinfo{journal}{Science}} \textbf{\bibinfo{volume}{365}},
  \bibinfo{pages}{565--570} (\bibinfo{year}{2019}).
\newblock \eprint{1906.11476}.

\bibitem{Ravi+2019}
\bibinfo{author}{{Ravi}, V.} \emph{et~al.}
\newblock \bibinfo{title}{{A fast radio burst localized to a massive galaxy}}.
\newblock \emph{\bibinfo{journal}{Nature}} \textbf{\bibinfo{volume}{572}},
  \bibinfo{pages}{352--354} (\bibinfo{year}{2019}).

\bibitem{Prochaska+2019}
\bibinfo{author}{{Prochaska}, J.~X.} \emph{et~al.}
\newblock \bibinfo{title}{{The low density and magnetization of a massive
  galaxy halo exposed by a fast radio burst}}.
\newblock \emph{\bibinfo{journal}{Science}} \textbf{\bibinfo{volume}{366}},
  \bibinfo{pages}{231--234} (\bibinfo{year}{2019}).
\newblock \eprint{1909.11681}.

\bibitem{Marcote+2020}
\bibinfo{author}{{Marcote}, B.} \emph{et~al.}
\newblock \bibinfo{title}{{A repeating fast radio burst source localized to a
  nearby spiral galaxy}}.
\newblock \emph{\bibinfo{journal}{Nature}} \textbf{\bibinfo{volume}{577}},
  \bibinfo{pages}{190--194} (\bibinfo{year}{2020}).
\newblock \eprint{2001.02222}.

\bibitem{Petroff+2016}
\bibinfo{author}{{Petroff}, E.} \emph{et~al.}
\newblock \bibinfo{title}{{FRBCAT: The Fast Radio Burst Catalogue}}.
\newblock \emph{\bibinfo{journal}{PASA}} \textbf{\bibinfo{volume}{33}},
  \bibinfo{pages}{e045} (\bibinfo{year}{2016}).
\newblock \eprint{1601.03547}.

\bibitem{Platts+2019}
\bibinfo{author}{{Platts}, E.} \emph{et~al.}
\newblock \bibinfo{title}{{A living theory catalogue for fast radio bursts}}.
\newblock \emph{\bibinfo{journal}{Physics Reports}}
  \textbf{\bibinfo{volume}{821}}, \bibinfo{pages}{1--27}
  (\bibinfo{year}{2019}).
\newblock \eprint{1810.05836}.

\bibitem{Pen+2015}
\bibinfo{author}{{Pen}, U.-L.} \& \bibinfo{author}{{Connor}, L.}
\newblock \bibinfo{title}{{Local Circumnuclear Magnetar Solution to
  Extragalactic Fast Radio Bursts}}.
\newblock \emph{\bibinfo{journal}{ApJ}} \textbf{\bibinfo{volume}{807}},
  \bibinfo{pages}{179} (\bibinfo{year}{2015}).
\newblock \eprint{1501.01341}.

\bibitem{Metzger+2017}
\bibinfo{author}{{Metzger}, B.~D.}, \bibinfo{author}{{Berger}, E.} \&
  \bibinfo{author}{{Margalit}, B.}
\newblock \bibinfo{title}{{Millisecond Magnetar Birth Connects FRB 121102 to
  Superluminous Supernovae and Long-duration Gamma-Ray Bursts}}.
\newblock \emph{\bibinfo{journal}{ApJ}} \textbf{\bibinfo{volume}{841}},
  \bibinfo{pages}{14} (\bibinfo{year}{2017}).
\newblock \eprint{1701.02370}.

\bibitem{Michilli+2018}
\bibinfo{author}{{Michilli}, D.} \emph{et~al.}
\newblock \bibinfo{title}{{An extreme magneto-ionic environment associated with
  the fast radio burst source FRB 121102}}.
\newblock \emph{\bibinfo{journal}{Nature}} \textbf{\bibinfo{volume}{553}},
  \bibinfo{pages}{182--185} (\bibinfo{year}{2018}).
\newblock \eprint{1801.03965}.

\bibitem{Margalit+2018}
\bibinfo{author}{{Margalit}, B.} \& \bibinfo{author}{{Metzger}, B.~D.}
\newblock \bibinfo{title}{{A Concordance Picture of FRB 121102 as a Flaring
  Magnetar Embedded in a Magnetized Ion-Electron Wind Nebula}}.
\newblock \emph{\bibinfo{journal}{ApJL}} \textbf{\bibinfo{volume}{868}},
  \bibinfo{pages}{L4} (\bibinfo{year}{2018}).
\newblock \eprint{1808.09969}.

\bibitem{Bochenek+2020a}
\bibinfo{author}{{Bochenek}, C.~D.} \emph{et~al.}
\newblock \bibinfo{title}{{STARE2: Detecting Fast Radio Bursts in the Milky
  Way}}.
\newblock \emph{\bibinfo{journal}{PASP}} \textbf{\bibinfo{volume}{132}},
  \bibinfo{pages}{034202} (\bibinfo{year}{2020}).
\newblock \eprint{2001.05077}.

\bibitem{Palmer+2020}
\bibinfo{author}{{Palmer}, D.~M.}
\newblock \bibinfo{title}{{A Forest of Bursts from SGR 1935+2154}}.
\newblock \emph{\bibinfo{journal}{The Astronomer's Telegram}}
  \textbf{\bibinfo{volume}{13675}} (\bibinfo{year}{2020}).

\bibitem{Younes+2020}
\bibinfo{author}{{Younes}, G.} \emph{et~al.}
\newblock \bibinfo{title}{{Burst forest from SGR 1935+2154 as detected with
  NICER}}.
\newblock \emph{\bibinfo{journal}{The Astronomer's Telegram}}
  \textbf{\bibinfo{volume}{13678}} (\bibinfo{year}{2020}).

\bibitem{Kennea+2020}
\bibinfo{author}{{Kennea}, J.~A.}, \bibinfo{author}{{Beardmore}, A.~P.},
  \bibinfo{author}{{Page}, K.~L.} \& \bibinfo{author}{{Palmer}, D.~M.}
\newblock \bibinfo{title}{{SGR 1935+2154: Swift detection of enhanced X-ray
  emission and dust scattered halo}}.
\newblock \emph{\bibinfo{journal}{The Astronomer's Telegram}}
  \textbf{\bibinfo{volume}{13679}} (\bibinfo{year}{2020}).

\bibitem{Scholz+2020}
\bibinfo{author}{{Scholz}, P.}
\newblock \bibinfo{title}{{A bright millisecond-timescale radio burst from the
  direction of the Galactic magnetar SGR 1935+2154}}.
\newblock \emph{\bibinfo{journal}{The Astronomer's Telegram}}
  \textbf{\bibinfo{volume}{13681}} (\bibinfo{year}{2020}).

\bibitem{Mereghetti+2020a}
\bibinfo{author}{{Mereghetti}, S.} \emph{et~al.}
\newblock \bibinfo{title}{{INTEGRAL IBIS and SPI-ACS detection of a hard X-ray
  counterpart of the radio burst from SGR 1935+2154}}.
\newblock \emph{\bibinfo{journal}{The Astronomer's Telegram}}
  \textbf{\bibinfo{volume}{13685}}, \bibinfo{pages}{1} (\bibinfo{year}{2020}).

\bibitem{Zhang+2020a}
\bibinfo{author}{{Zhang}, S.-N.} \emph{et~al.}
\newblock \bibinfo{title}{{Insight-HXMT X-ray and hard X-ray detection of the
  double peaks of the Fast Radio Burst from SGR 1935+2154}}.
\newblock \emph{\bibinfo{journal}{The Astronomer's Telegram}}
  \textbf{\bibinfo{volume}{13696}} (\bibinfo{year}{2020}).

\bibitem{Kothes+2018}
\bibinfo{author}{{Kothes}, R.}, \bibinfo{author}{{Sun}, X.},
  \bibinfo{author}{{Gaensler}, B.} \& \bibinfo{author}{{Reich}, W.}
\newblock \bibinfo{title}{{A Radio Continuum and Polarization Study of SNR
  G57.2+0.8 Associated with Magnetar SGR 1935+2154}}.
\newblock \emph{\bibinfo{journal}{ApJ}} \textbf{\bibinfo{volume}{852}},
  \bibinfo{pages}{54} (\bibinfo{year}{2018}).
\newblock \eprint{1711.11146}.

\bibitem{Ranasinghe+2018}
\bibinfo{author}{{Ranasinghe}, S.}, \bibinfo{author}{{Leahy}, D.~A.} \&
  \bibinfo{author}{{Tian}, W.}
\newblock \bibinfo{title}{{New Distances to Four Supernova Remnants}}.
\newblock \emph{\bibinfo{journal}{Open Physics Journal}}
  \textbf{\bibinfo{volume}{4}}, \bibinfo{pages}{1--13} (\bibinfo{year}{2018}).
\newblock \eprint{1712.04515}.

\bibitem{Zhou+2020}
\bibinfo{author}{{Zhou}, P.} \emph{et~al.}
\newblock \bibinfo{title}{{Revisiting the distance, environment and supernova
  properties of SNR G57.2+0.8 that hosts SGR 1935+2154}}.
\newblock \emph{\bibinfo{journal}{arXiv e-prints}}
  \bibinfo{pages}{arXiv:2005.03517} (\bibinfo{year}{2020}).
\newblock \eprint{2005.03517}.

\bibitem{Keane2018}
\bibinfo{author}{{Keane}, E.~F.}
\newblock \bibinfo{title}{{The future of fast radio burst science}}.
\newblock \emph{\bibinfo{journal}{Nature Astronomy}}
  \textbf{\bibinfo{volume}{2}}, \bibinfo{pages}{865--872}
  (\bibinfo{year}{2018}).
\newblock \eprint{1811.00899}.

\bibitem{CHIME+2019}
\bibinfo{author}{{CHIME/FRB Collaboration}} \emph{et~al.}
\newblock \bibinfo{title}{{CHIME/FRB Discovery of Eight New Repeating Fast
  Radio Burst Sources}}.
\newblock \emph{\bibinfo{journal}{ApJL}} \textbf{\bibinfo{volume}{885}},
  \bibinfo{pages}{L24} (\bibinfo{year}{2019}).
\newblock \eprint{1908.03507}.

\bibitem{CHIME+2020}
\bibinfo{author}{{The CHIME/FRB Collaboration}} \emph{et~al.}
\newblock \bibinfo{title}{{Periodic activity from a fast radio burst source}}.
\newblock \emph{\bibinfo{journal}{arXiv e-prints}}
  \bibinfo{pages}{arXiv:2001.10275} (\bibinfo{year}{2020}).
\newblock \eprint{2001.10275}.

\bibitem{Chawla+2020}
\bibinfo{author}{{Chawla}, P.} \emph{et~al.}
\newblock \bibinfo{title}{{Detection of Repeating FRB 180916.J0158+65 Down to
  Frequencies of 300 MHz}}.
\newblock \emph{\bibinfo{journal}{arXiv e-prints}}
  \bibinfo{pages}{arXiv:2004.02862} (\bibinfo{year}{2020}).
\newblock \eprint{2004.02862}.

\bibitem{Cordes+2017}
\bibinfo{author}{{Cordes}, J.~M.} \emph{et~al.}
\newblock \bibinfo{title}{{Lensing of Fast Radio Bursts by Plasma Structures in
  Host Galaxies}}.
\newblock \emph{\bibinfo{journal}{ApJ}} \textbf{\bibinfo{volume}{842}},
  \bibinfo{pages}{35} (\bibinfo{year}{2017}).
\newblock \eprint{1703.06580}.

\bibitem{Gajjar+2018}
\bibinfo{author}{{Gajjar}, V.} \emph{et~al.}
\newblock \bibinfo{title}{{Highest Frequency Detection of FRB 121102 at 4-8 GHz
  Using the Breakthrough Listen Digital Backend at the Green Bank Telescope}}.
\newblock \emph{\bibinfo{journal}{ApJ}} \textbf{\bibinfo{volume}{863}},
  \bibinfo{pages}{2} (\bibinfo{year}{2018}).
\newblock \eprint{1804.04101}.

\bibitem{Hessels+2019}
\bibinfo{author}{{Hessels}, J.~W.~T.} \emph{et~al.}
\newblock \bibinfo{title}{{FRB 121102 Bursts Show Complex Time-Frequency
  Structure}}.
\newblock \emph{\bibinfo{journal}{ApJL}} \textbf{\bibinfo{volume}{876}},
  \bibinfo{pages}{L23} (\bibinfo{year}{2019}).
\newblock \eprint{1811.10748}.

\bibitem{Majid+2020}
\bibinfo{author}{{Majid}, W.~A.} \emph{et~al.}
\newblock \bibinfo{title}{{A Dual-Band Radio Observation of FRB 121102 with the
  Deep Space Network and the Detection of Multiple Bursts}}.
\newblock \emph{\bibinfo{journal}{arXiv e-prints}}
  \bibinfo{pages}{arXiv:2004.06845} (\bibinfo{year}{2020}).
\newblock \eprint{2004.06845}.

\bibitem{Spitler+2016}
\bibinfo{author}{{Spitler}, L.~G.} \emph{et~al.}
\newblock \bibinfo{title}{{A repeating fast radio burst}}.
\newblock \emph{\bibinfo{journal}{Nature}} \textbf{\bibinfo{volume}{531}},
  \bibinfo{pages}{202--205} (\bibinfo{year}{2016}).
\newblock \eprint{1603.00581}.

\bibitem{Law+2017}
\bibinfo{author}{{Law}, C.~J.} \emph{et~al.}
\newblock \bibinfo{title}{{A Multi-telescope Campaign on FRB 121102:
  Implications for the FRB Population}}.
\newblock \emph{\bibinfo{journal}{ApJ}} \textbf{\bibinfo{volume}{850}},
  \bibinfo{pages}{76} (\bibinfo{year}{2017}).
\newblock \eprint{1705.07553}.

\bibitem{Zhang+2020b}
\bibinfo{author}{{Zhang}, C.~F.} \emph{et~al.}
\newblock \bibinfo{title}{{A highly polarised radio burst detected from SGR
  1935+2154 by FAST}}.
\newblock \emph{\bibinfo{journal}{The Astronomer's Telegram}}
  \textbf{\bibinfo{volume}{13699}} (\bibinfo{year}{2020}).

\bibitem{Lyutikov2002}
\bibinfo{author}{{Lyutikov}, M.}
\newblock \bibinfo{title}{{Radio Emission from Magnetars}}.
\newblock \emph{\bibinfo{journal}{ApJL}} \textbf{\bibinfo{volume}{580}},
  \bibinfo{pages}{L65--L68} (\bibinfo{year}{2002}).
\newblock \eprint{astro-ph/0206439}.

\bibitem{Lyutikov+2020}
\bibinfo{author}{{Lyutikov}, M.} \& \bibinfo{author}{{Popov}, S.}
\newblock \bibinfo{title}{{Fast Radio Bursts from reconnection events in
  magnetar magnetospheres}}.
\newblock \emph{\bibinfo{journal}{arXiv e-prints}}
  \bibinfo{pages}{arXiv:2005.05093} (\bibinfo{year}{2020}).
\newblock \eprint{2005.05093}.

\end{thebibliography}
\end{document}